\documentclass[sn-mathphys,onecolumn]{sn-jnl}

\usepackage{graphicx}%
\usepackage{multirow}%
\usepackage{setspace}
\usepackage{amsmath,amssymb,amsfonts}%
\usepackage{amsthm}%
\usepackage{mathrsfs}%
\usepackage[title]{appendix}%
\usepackage{xcolor}%
\usepackage{textcomp}%
\usepackage{manyfoot}%
\usepackage{booktabs}%
\usepackage{algorithm}%
\usepackage{algorithmicx}%
\usepackage{algpseudocode}%
\usepackage{listings}%
\usepackage{hyperref}
\usepackage{cleveref}
\usepackage{placeins}
\usepackage{siunitx}
\newcommand{\supplementarysection}{
  \setcounter{figure}{0}% Reset figure counter
  \setcounter{table}{0}% Reset table counter
  \setcounter{section}{1}
  \renewcommand{\thefigure}{S\arabic{figure}}% Figure numbering as S1, S2, ...
  \renewcommand{\thetable}{S\arabic{table}}% Table numbering as S1, S2, ...
  \section*{Supplementary Information}% Numbered section title
  \addcontentsline{toc}{section}{Supplementary Information}% Add to TOC
}

\theoremstyle{thmstyleone}%
\theoremstyle{thmstyletwo}%

\theoremstyle{thmstylethree}%

\begin{document}
\onehalfspacing

\title[Article Title]{Organic Hydrogen Sensors for Potential Use in Safety-Critical Environments}


\author[1]{\fnm{Annika} \sur{Morgenstern}}

\author[2]{\fnm{Lucas} \sur{Viriato}}

\author[3]{\fnm{Frank} \sur{Ortmann}}

\author[4]{\fnm{Christopher} \sur{Bickmann}}

\author[1,5]{\fnm{Lukas} \sur{Hertling}}

\author[1,6]{\fnm{Dominik} \sur{Weber}}

\author[1,5]{\fnm{Dietrich R.T.} \sur{Zahn}}

\author[4]{\fnm{Karla} \sur{Hiller}}

\author[2]{\fnm{Thomas} \sur{von Unwerth}}

\author[6]{\fnm{Daniel} \sur{Schondelmaier}}

\author*[1,5]{\fnm{Georgeta} \sur{Salvan}}

\affil*[1]{\orgdiv{Institute of Physics}, \orgaddress{\street{Reichenhainer Str. 70}, \city{Chemnitz}, \postcode{09126}, \state{Germany}}}

\affil[2]{\orgdiv{Institute of Mechanical Engineering}, \orgaddress{\street{Reichenhainer Str. 70}, \city{Chemnitz}, \postcode{09126}, \state{Germany}}}

\affil[3]{\orgdiv{Department of Chemistry, TUM School of Natural Sciences and Atomistic Modeling Center, Munich Data Science Institute}, \orgaddress{\street{Lichtenbergstr. 4}, \city{Munich}, \postcode{85748}, \state{Germany}}}

\affil[4]{\orgdiv{Center for Micro and Nano Technologies}, \orgaddress{\street{{Reichenhainer Str. 70}, \city{Chemnitz}, \postcode{09126}, \state{Chemnitz}}}}

\affil[5]{\orgdiv{Research Center for Materials, Architectures and Integration of Nanomembranes (MAIN)}, \orgaddress{\street{Reichenhainer Str. 70}, \city{Chemnitz}, \postcode{09126}, \state{Germany}}}

\affil[6]{\orgdiv{Institute of Physical Engineering/Computer Science}, \orgaddress{\street{Peter-Breuer Straße 2}, \city{Zwickau}, \postcode{08056}, \state{Zwickau}, \country{Germany}}}


\abstract{
\unboldmath
Accurate monitoring of the hydrogen concentration is critical for optimizing fuel cell performance, minimizing purge losses, and reducing long-term degradation. Conventional hydrogen sensors often rely on catalytic materials and face limitations such as the need of oxygen purging when operated in fuel cell environments. Here, we report the discovery of a novel hydrogen-sensing mechanism based on organic molecules, without the use of catalytic metals. The sensor is based on a typical vertical stack geometry, containing $\mathrm{Alq_3}$ as active organic material. Upon exposure to hydrogen, the device shows an increase in resistivity, yielding a reliable sensor signal that varies linearly with hydrogen concentration, temperature, and humidity, and exhibits a relative response of up to $3.5 \, \%$ at $100 \, \% \mathrm{vol}$ hydrogen. By exposing the sensor to an external magnetic field, the rise and fall times of the sensor response were found to be tunable. This novel organic sensor demonstrates sensitivity across a wide range of hydrogen concentrations under fuel cell-relevant conditions and beyond. This new class of hydrogen
sensors with high miniaturization potential and cost efficiency paves the way for real-time hydrogen monitoring and advanced control strategies in fuel cells, the chemical industry, or energy storage applications.}

%%================================%%
%% Sample for structured abstract %%
%%================================%%

\keywords{ Organic Molecules, Hydrogen Sensor, Fuel Cell Application, Organic Magnetic Field Effects}
\maketitle

\clearpage
\newpage
\section{Introduction}\label{sec1}

Hydrogen is widely regarded as a key energy carrier in the transition toward a decarbonized economy, playing a central role in fuel cell technology, the chemical industry, and energy storage systems \cite{moran2019pem, en17205158}. Accurate, rapid, and reliable hydrogen sensing is essential for ensuring operational safety, maximizing efficiency, and enabling robust system integration. Consequently, there is a strong demand for sensor systems able to detect hydrogen under dynamic and application-relevant conditions.

Most of the existing hydrogen sensor technologies are primarily optimized for safety applications, focusing on leak detection in the low concentration range, from a few ppm up to approximately $4\mathrm{\% \, vol}$. For this purpose, typical sensor concepts include the use of catalytic materials \cite{awang2014gas, lee2011micromachined, honeywell_705HT}, metal-oxides \cite{ji2019gas}, and electrochemical-based sensors \cite{hubert2011hydrogen}. However, these concepts typically require the presence of oxygen to regenerate the sensing layer, making them unsuitable for use in oxygen-depleted environments, such as in closed-loop hydrogen recirculation systems of fuel cells \cite{tian2020hydrogen}. A detailed description of the functionality and requirements for sensors in fuel cell systems can be found in the Supplementary Information (cf. Figure S1).

An established class of hydrogen sensors is based on gas-sensitive metals such as palladium or platinum \cite{koo2020chemiresistive}, which are already commercially available \cite{honeywell_705HT} (detection range $\mathrm{0 - 4 \, \% vol}$). While these materials offer high sensitivity and selectivity, they tend to become brittle after prolonged hydrogen exposure, requiring complex encapsulation strategies. Additionally, the use of noble metals imply high fabrication costs.

To address some of these limitations, Mandal \textit{et al.} \cite{mandal2025robust} recently proposed a hybrid sensor concept that combines organic materials with a catalytic layer containing Pt. The use of organic semiconductors significantly improved the sensitivity and response time of the device. The sensor operates on the principle of physisorption: oxygen molecules diffuse through the organic layer and adsorb onto the underlying Pt electrode, where they are catalytically activated. This leads to an increase in device current due to oxygen-induced doping of the Pt layer. When hydrogen is introduced, it diffuses through the organic material and dissociates at the platinum (Pt) surface into atomic hydrogen. These atoms react with the adsorbed oxygen to form water, which reduces the doping effect and results in an increase in the device resistance. This type of sensor requires cyclic regeneration by purging with oxygen and has only been tested at hydrogen concentrations up to a few ppm. It is therefore unsuitable for use in fuel cell systems, where concentrations close to $100\, \% \mathrm{vol}$ $\mathrm{H_2}$ must be detected, and oxygen purging is not feasible.

In this study, we propose an innovative, highly responsive sensor system based on organic molecules without the need for oxygen regeneration or catalytic metals \cite{mandal2025robust, en17205158}. Given the requirements of fuel cell applications, the sensor was designed to operate reliably over a wide concentration range, even in a pure hydrogen atmosphere. The application of an external magnetic field modulated the rise and fall times of the sensor. In addition, no significant cross-sensitivity to ethanol, toluene, or air was detected. Considering the low production costs, biocompatibility, geometric flexibility \cite{yu2005spin}, and the obtained strong sensor signal, we propose that organic materials are highly suited for further $\mathrm{H_2}$ sensing investigations and have a strong potential for integration into fuel cell systems, and beyond.

\section{Results}\label{sec2}
\subsection{Sensor Geometries and Responsivity towards Hydrogen }
\begin{figure*}
    \centering
    \includegraphics[width=1\linewidth]{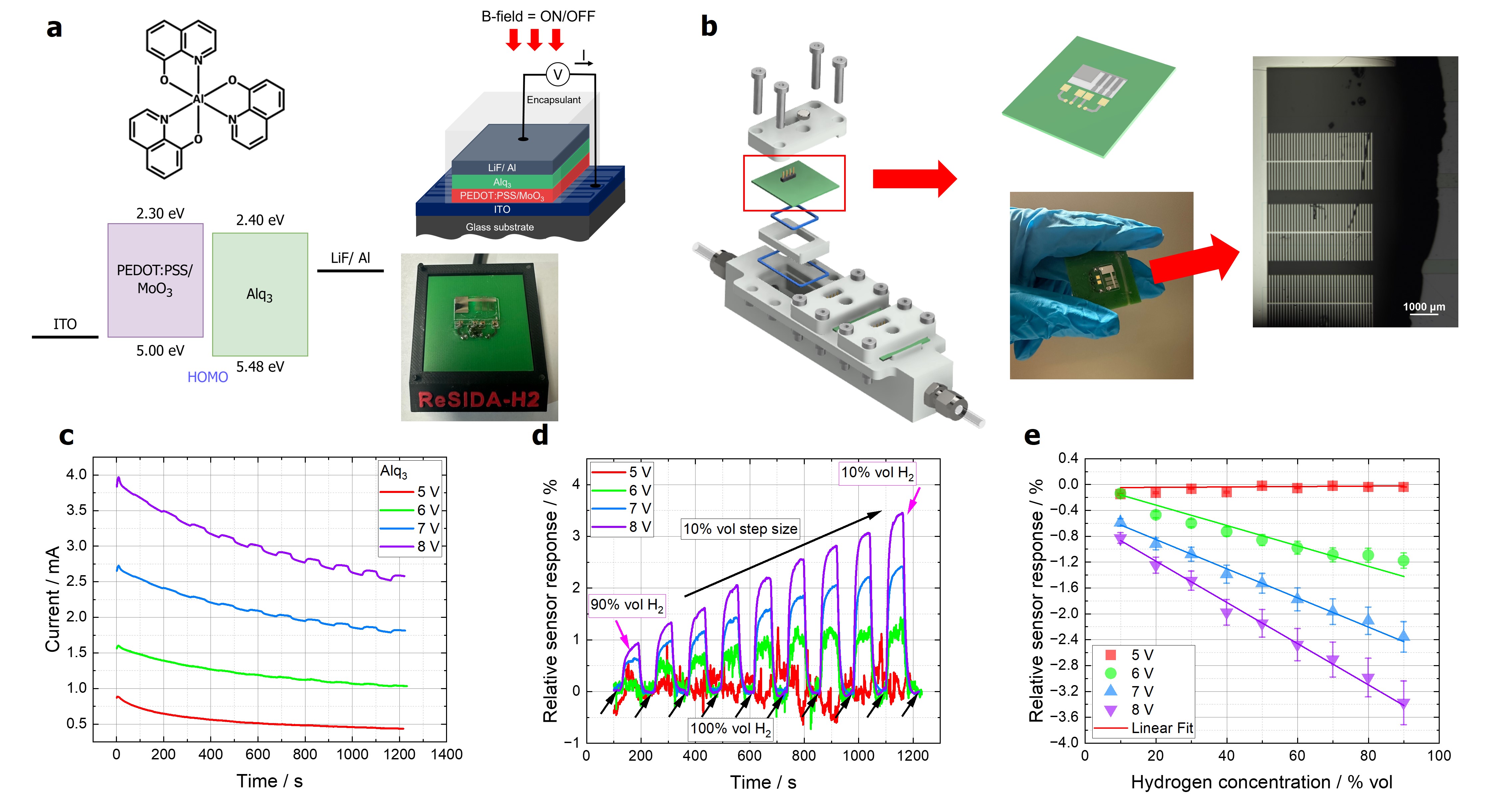}
    \caption{\textbf{a} Chemical structure of the hydrogen-sensitive organic compound $\mathrm{Alq_3}$. Additionally, the band alignment for the $\mathrm{Alq_3}$-based $\mathrm{H_2}$ sensor is shown. \textbf{b} Demonstrator chamber together with the used sensor, placed on a printed circuit board (PCB). On the right-hand side, an optical microscopy image of the full layer stack after the deposition of the epoxy resin is shown. The structured electrode with a width of the bars of \SI{50}{\micro m} can be seen. \textbf{c} Raw sensor response for various bias voltages and hydrogen concentrations. \textbf{d} Sensor response corrected by the background signal using a spline function (IRSQR). \textbf{e} Linear behavior for the sensor response towards hydrogen concentration.}
    \label{fig:layer-stack-scheme}
\end{figure*}

The sensor structure is based on a vertical stack geometry, with Tris(8-hydroxyquinolinato)aluminium ($\mathrm{Alq_3}$, cf. Figure \ref{fig:layer-stack-scheme} \textbf{a}) as an active material. $\mathrm{Alq_3}$ is a small molecule commonly used in light-emitting applications, like organic light-emitting diodes \cite{debsharma2022recent, kwong2005efficiency}. Additionally, a hole-transport layer (PEDOT:PSS/$\mathrm{MoO_3}$) was introduced to improve the band alignment (cf. Figure \ref{fig:layer-stack-scheme} \textbf{a}). The lowest unoccupied molecular orbital (LUMO) and the highest occupied molecular orbital (HOMO) energy values for $\mathrm{Alq_3}$ are \SI{2.40}{eV} and \SI{5.48}{eV} \cite{tolkki2012organometallic}, respectively. In contrast to a typical vertical structure device (\textit{e.g.} organic light emitting diode), the top electrode was microstructured (cf. Figure \ref{fig:layer-stack-scheme} \textbf{b}) to enable a direct exposure area of the organic material to the hydrogen gas. The mask layout, used for evaporating both lithium fluoride (LiF) and aluminium (Al), is depicted in Figure S2. The distance between the bars as well as the bar width were chosen to be \SI{50}{\micro m}. LiF is typically employed as an electron injection layer to reduce the work function of the aluminum cathode, thereby enhancing the energy level alignment between the organic layer and the top electrode. This leads to more efficient electron injection, resulting in improved device efficiency and a lower turn-on voltage \cite{brown2000lif, turak2021role}. To prevent atmospheric-induced degradation, the sample was encapsulated with an epoxy resin via blade coating. Additionally, the effects of various encapsulation strategies on the sensor signal were compared, \textit{e.g.} encapsulation by epoxy resin with and without an additional microscope slide. The resin is semipermeable to hydrogen, enabling selective gas permeation \cite{gajda2021hydrogen}, especially at thicknesses in the mm region. The thickness of the epoxy encapsulation layer, as determined by profilometry, was $(0.159 \pm 0.059) \, \mathrm{mm}$ for our devices. Within the test chamber, which is depicted in Figure \ref{fig:layer-stack-scheme} \textbf{b}, at least two sensors were measured simultaneously. The chamber was connected to a gas mixing system, which is schematically shown in Figure S3. The system enables the mixing of multiple gases and analytes, including evaporated solutions (\textit{e.g.}, ethanol), and allows for water evaporation to achieve humidity levels of up to $100 \ \%$ relative humidity (r.h.). Furthermore, a climate chamber was used to perform temperature-dependent measurements up to $80 ^\circ$C. Unless stated otherwise, all measurements were performed at room temperature ($23.5 ^\circ$C, $0 \ \% \ \mathrm{r.h.}$).
\FloatBarrier
\begin{figure*}
    \centering
    \includegraphics[width=1\linewidth]{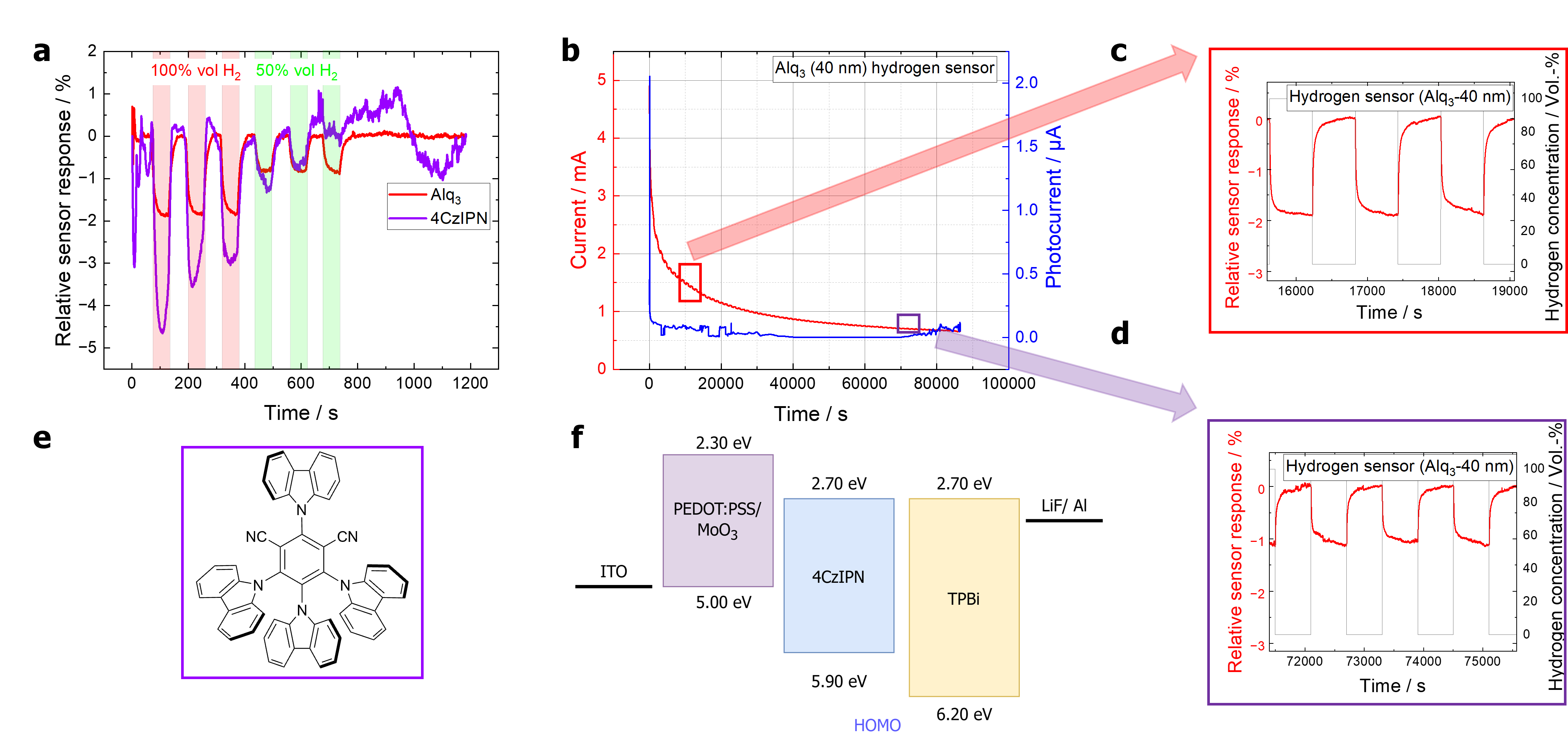}
    \caption{\textbf{a} Sensing behavior of $\mathrm{Alq_3}$ compared to the common conor-acceptor type emitter molecule 4CzIPN. $\mathrm{Alq_3}$ exhibits a much more stable response to the recurring hydrogen concentration. \textbf{b} Current and photocurrent response of an $\mathrm{Alq_3}$ based sensor with the marked regions shown in \textbf{c}. In \textbf{c} and \textbf{d}, the background was subtracted using a spline function. Even without radiative recombination, after the photcurrent drop, the sensor response remains stable. \textbf{e} Molecular structure for the 4CzIPN molecule, with the corresponding band alignment in \textbf{f}.}
    \label{fig:comparison 4czIPN-Alq3-comp-photocurrent}
\end{figure*}

The current through the sensor device was monitored for different applied bias voltages in the vicinity of the turn-on voltage, ranging from \SI{5}{V} to \SI{8}{V}. The turn-on voltage corresponds to the minimum applied voltage at which the device begins to emit detectable light, marking the onset of efficient charge injection and recombination in the organic layers. As the $\mathrm{Alq_3}$-containing sensors emit light, the light intensity was monitored using a photodiode. The turn-on voltage was found to be $V_\mathrm{on} = (7.21 \pm 0.72)\ \mathrm{V}$ (cf. Figure S4 \textbf{a}), which was determined at a luminance of $10 \, \mathrm{cd/m^2}$. This relatively large value (compared to literature $V_\mathrm{on} = 5 \, \mathrm{V}$ \cite{mohd2015electroluminescence}) can be attributed to the choice of the layer stack and, in particular, to the limited overlap between the structured top electrode and the bottom electrode, which was not optimized for Organic-light emitting diode (OLED) efficiency but for hydrogen sensing instead. As a result, the effective volume of organic material available for charge transport is reduced, which restricts current pathways and increases the turn-on voltage. The sensor response measure corresponds to the relative current difference depending on the hydrogen concentration, namely $\Delta I / I_0$. To better observe the response dynamics, the hydrogen concentration was reduced by $10\mathrm{\% \, vol}$ after \SI{120}{s} and restored to $100\mathrm{\% \, vol}$ at \SI{60}{s}. As evident from Figure \ref{fig:layer-stack-scheme} \textbf{c}, the devices degrade over time. The natural current evolution with time for vertical stack organic devices is commonly known to follow an exponential decay of the form $I(t) = I_0 \mathrm{exp}^{(-t/\tau)} + a$, where $\mathrm{\tau}$ is the decay lifetime, and $a$ the baseline given by the detector noise level \cite{kohler2015electronic}. 
However, in non-optimized devices, the decay may deviate from this expected trend. In our case, in the measured signal, the natural exponential decay of the current $I(t)$ is superimposed by a response to the hydrogen concentration, denoted as $I(t, \mathrm{c_{gas}})$. To isolate the hydrogen-related contribution, a spline fit function (Iterative Reweighted Spline Quantile Regression (IRSQR) from the pybaselines library \cite{rapp_pybaselines}) was applied to the raw measurement data. Subsequently, the baseline, represented as $I(t)$, was subtracted from the raw measurement data. A detailed example is shown in the SI in Figure S4. The function $I\mathrm{(c_{gas}}) = I\mathrm{(t, c_{gas}})-I\mathrm{(t)}$ will be called "sensor response" in the following. As can be seen from Figure \ref{fig:layer-stack-scheme} \textbf{d/e}, the sensor response increases with increasing bias voltage. Furthermore, a linear correlation was observed between the hydrogen concentration difference, \textit{i.e.} the difference between a given hydrogen concentration and the "zero" line, defined as $100\mathrm{\% \, vol} \ \mathrm{H_2}$), and the sensor signal, indicating the suitability of the $\mathrm{Alq_3}$-based device as a hydrogen sensing element. At a bias voltage of \SI{5}{V}\textit{i.e.}, below the turn-on voltage, the sensor exhibits a high signal-to-noise ratio. Increasing the bias voltage leads to an increased sensor signal, with a current response in the \si{\micro A} range, without any additional amplification. To optimize the trade-off between sensor performance and operational stability, a constant bias of \SI{7}{V} was applied in all subsequent measurements. This choice is further supported by the J-V measurements provided in Figure S5 \textbf{a}, where \SI{7}{V} corresponds to a stable regime and a power consumption of $\simeq 28$ mW. 

\subsection{Mechanistic Insights Into Hydrogen Sensing in Organic Devices}
To access whether the hydrogen response is intrinsically linked to the metal complex $\mathrm{Alq_3}$ or can also arise in a metal-free organic semiconductor, we compared $\mathrm{Alq_3}$ with the donor–acceptor molecule 1,2,3,5-tetrakis(carbazol-9-yl)-4,6-dicyanobenzene (4CzIPN) \cite{uoyama2012highly, streiter2020impact} which does not contain a metal center. This study tested whether the metal complex in $\mathrm{Alq_3}$ plays a decisive role in the sensing mechanism, as 4CzIPN does not contain a metal complex. The molecular structure of 4CzIPN is shown in Fig. \ref{fig:comparison 4czIPN-Alq3-comp-photocurrent}\textbf{e}, and the corresponding band alignment of the device based on 4CzIPN is illustrated in Fig. \ref{fig:comparison 4czIPN-Alq3-comp-photocurrent}\textbf{f}. The sensor response of the 4CzIPN-based device was directly compared to that of an $\mathrm{Alq_3}$-based device, with the respective signals shown in Fig. \ref{fig:comparison 4czIPN-Alq3-comp-photocurrent}\textbf{a}. Details regarding the optoelectronic characterization of 4CzIPN can be found elsewhere \cite{morgenstern2025unlocking}. The sensor based on 4CzIPN also exhibits sensitivity to hydrogen exposure, showing an even stronger response at the first cycle compared to the $\mathrm{Alq_3}$ device. Nevertheless, the signal-to-noise ratio appears to be much higher, related to the higher absolute current in the device (cf. Figure S6).  However, the sensor response of 4CzIPN devices strongly degrades, which limits their suitability for integration into long-term measurement systems. We assume that the observed behavior may be related to differences in the molecular rigidity and the intermolecular interactions between the two molecular systems 4CzIPN and $\mathrm{Alq_3}$. As previously reported, molecular rigidity significantly influences charge transport properties \cite{godi2024enhancing}. These findings indicate that a metal center is not required for the hydrogen response itself, while molecular rigidity and intermolecular interactions critically influence its stability. Hence, in the following, we will focus on the layer stack containing $\mathrm{Alq_3}$ as active material.

As the $\mathrm{Alq_3}$-based devices emit light in the visible region, the current and photocurrent were monitored over a period of 24 hours, during which hydrogen was cyclically turned on ($100\mathrm{\% \, vol}\ \mathrm{H_2}$) and off ($100\mathrm{\% \, vol} \ \mathrm{N_2}$) every 10 minutes. To measure the photocurrent, the light emitted by the sensor was monitored using an additional photodiode positioned in front of the emitting region. The corresponding results are presented in Figure \ref{fig:comparison 4czIPN-Alq3-comp-photocurrent} \textbf{b}, with detailed views in the insets of Figure \ref{fig:comparison 4czIPN-Alq3-comp-photocurrent} \textbf{c/d}. As shown, the photocurrent rapidly decreases within a few minutes. Notably, despite the decline in photocurrent, the electrical current remains at a low signal-to-noise ratio, and the sensor retains its sensitivity to hydrogen. This indicates that radiative recombination is not required to generate a sensor response to $\mathrm{H_2}$ exposure. Thus, the sensor performance does not require light outcoupling. This finding opens up a wide range of alternative device configurations, such as organic field-effect transistors, for future investigations. Further, alternative molecules that do not emit light in the visible region can be explored.

\begin{figure*}
    \centering
    \includegraphics[width=1\linewidth]{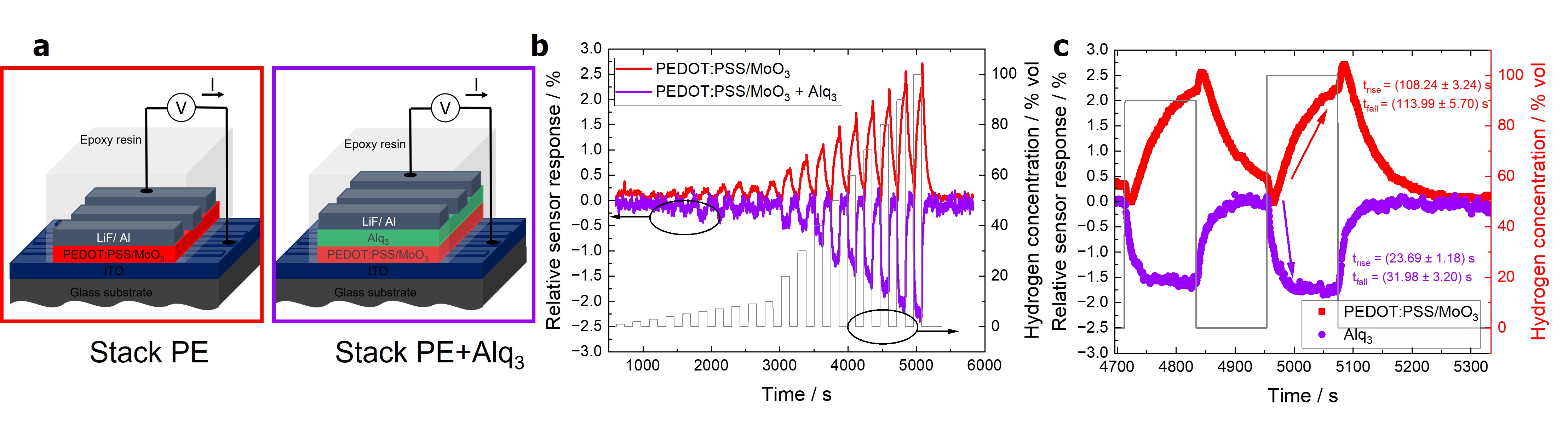}
    \caption{   \textbf{a} Schematic of the two investigated devices: Stack PE and Stack PE+$\mathrm{Alq_3}$. \textbf{b} Response of both devices to hydrogen exposure: the resistance decreases for the Stack PE, while it increases for the  Stack PE+$\mathrm{Alq_3}$. \textbf{c} Rise and fall times of both devices, showing that the  Stack PE exhibits nearly five times longer rise and fall times than the  Stack PE+$\mathrm{Alq_3}$.}
    \label{fig:PEDOT-Alq3-comparison}
\end{figure*}

In order to understand the role of $\mathrm{Alq_3}$ in the $\mathrm{H_2}$ sensing, several other device architectures were tested. In a first step, devices without the $\mathrm{Alq_3}$ layer, \textit{i.e.}, containing only the hole-transport layer, $\mathrm{PEDOT:PSS/MoO_3}$ (in the following named Stack PE), were investigated and compared to the full layer stack containing $\mathrm{Alq_3}$ (in the following named Stack PE+$\mathrm{Alq_3}$), schematically shown in Figure \ref{fig:PEDOT-Alq3-comparison} \textbf{a}. Interestingly, the hydrogen response exhibited an opposite sign when measuring the Stack PE compared to the Stack PE+$\mathrm{Alq_3}$ (see Figure \ref{fig:PEDOT-Alq3-comparison} \textbf{b}). While the resistance increases for the Stack PE+$\mathrm{Alq_3}$, it decreases for the Stack PE under hydrogen exposure. This indicates that, in the Stack PE+$\mathrm{Alq_3}$, the effect of $\mathrm{PEDOT:PSS/MoO_3}$ must be compensated, suggesting that the intrinsic response of $\mathrm{Alq_3}$ is likely even stronger. Moreover, a comparison of the line shape and thus the rise times revealed a pronounced difference: the Stack PE exhibited a rise time nearly five times larger than the one obtained for the Stack PE+$\mathrm{Alq_3}$, and the same applies to the fall time (cf. Figure \ref{fig:PEDOT-Alq3-comparison} \textbf{c}).
Both the $\mathrm{Alq_3}$ layer and the underlying layer contribute to the overall sensor response. The reduced rise and fall times in the Stack PE+$\mathrm{Alq_3}$ compared to the Stack PE indicate a bulk effect. The observed reversal in the sign of the sensor response suggests an interface-controlled sensing mechanism arising from modulation of the charge-injection barrier, providing direct evidence that interfacial barrier modulation governs the observed behavior.
In addition to the $\mathrm{PEDOT:PSS/MoO_3}$ layer, the possible contribution of the electrodes was examined, since ITO has been shown to exhibit sensitivity towards hydrogen in previous studies \cite{almaev2022ito}. However, a strong cross-sensitivity was observed, as well as a saturation at low hydrogen gas concentrations, unsuitable for deployment in fuel cell systems. Furthermore, ITO films had to be activated, by \textit{e.g.} annealing \cite{almaev2022ito}. The response of Al to hydrogen is expected to be negligible \cite{zhang2012hydrogen}, whereas for LiF, no reliable experimental data are currently available.

Two separate layers of ITO and LiF, together with Al, were fabricated to replicate the bottom and top electrodes. The lateral resistance of the ITO layer was measured under hydrogen exposure, using the same ITO as in the device stack. In addition, a \SI{0.8}{nm} LiF layer and a \SI{110}{nm} Al layer were deposited on a microscope slide under the same conditions as used for device fabrication. As evident from Figure S7, no significant response towards the hydrogen gas was monitored for either electrode, excluding a bulk contribution of the electrodes to the observed sensor signal. \newline
Nonetheless, previous studies showed that metal oxides act as catalysts to dissociate molecular hydrogen \cite{almaev2022ito}. In all reported studies, this chemical reaction required oxygen and resulted in the formation of water and free electrons. 
\begin{equation}
    \mathrm{H_2 + O^- (adsorbed) \rightarrow H_2O + e^-}
\end{equation}
If such a reaction would take place in the Stack PE+$\mathrm{Alq_3}$ devices, it would result in water formation, which is critical, as $\mathrm{Alq_3}$ is highly sensitive to moisture, leading to gradual degradation of device performance over a short time. Aziz \textit{et al.} \cite{aziz1998humidity} systematically investigated the effect of humidity on $\mathrm{Alq_3}$ molecules, confirming their susceptibility to moisture-induced degradation. It was found that a crystallization of the $\mathrm{Alq_3}$ occurred upon humidity exposure. In order to check for an eventual humidity-induced crystallization, microscopy images of the devices were taken before and after hydrogen exposure (see Figure S8) at three different magnifications. The microscopy images show neither crystallization nor morphological differences. Thus, a chemical reaction producing water can be excluded.

\begin{figure*}
    \centering
    \includegraphics[width=0.7\linewidth]{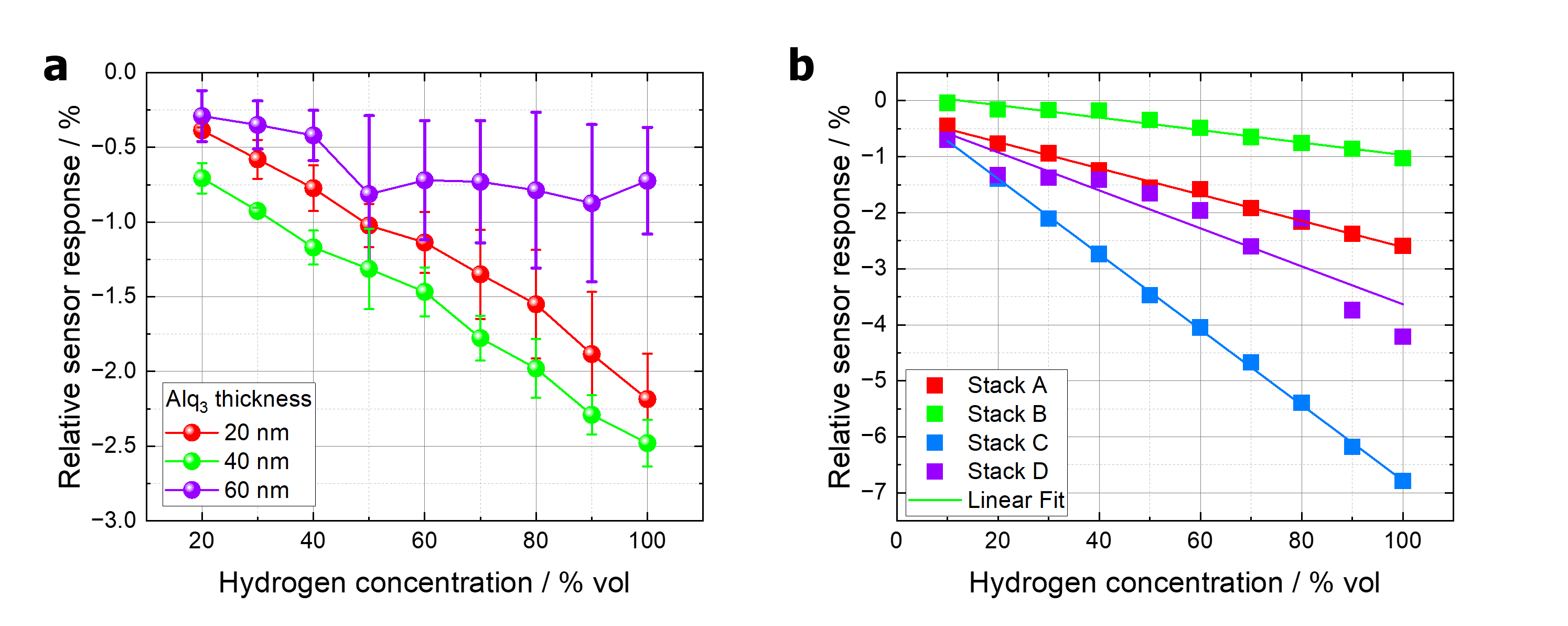}
    \caption{\textbf{a} Impact onto the sensor response of the active layer thickness of the $\mathrm{Alq_3}$ layer, which was varied from \SI{20}{nm} to \SI{60}{nm}, in \SI{20}{nm} steps, and \textbf{b} shows the difference in sensor response upon various encapsulation methods, choosing either a structured or closed electrode, and encapsulation either by epoxy resin with or without a microscope slide (cf. Figure S8).}
    \label{fig:thickness-variation-and-encapsulation}
\end{figure*}

To investigate the bulk $\mathrm{Alq_3}$ contribution to the sensor response,   Stack PE+$\mathrm{Alq_3}$ with three different $\mathrm{Alq_3}$ thicknesses were fabricated. The results are shown in Figure \ref{fig:thickness-variation-and-encapsulation} \textbf{a}. An increase of the sensor response towards increasing hydrogen gas concentrations can be seen for \SI{20}{nm} and \SI{40}{nm} thickness of $\mathrm{Alq_3}$, and the relative sensor response is increased when comparing \SI{20}{nm} and \SI{40}{nm}. By further increasing the $\mathrm{Alq_3}$ thickness to \SI{60}{nm}, a saturation of the effect was found. At this point, increasing the hydrogen concentration does not significantly impact the sensor response. Thus, a bulk contribution of $\mathrm{Alq_3}$ is indeed present. To get a deeper insight in the role of $\mathrm{Alq_3}$ in the sensor response, the interaction between $\mathrm{H_2}$ and $\mathrm{Alq_3}$ molecules was investigated by quantum chemical calculations. Details can be found in the section \ref{Sensor Principle}.

To evaluate the influence of the encapsulation and the structured electrode, four different stacks were investigated, namely structured electrode (cf. Figure  S2) + epoxy resin (Stack A), structured electrode + epoxy resin + microscope slide (Stack B), closed electrode + epoxy resin (Stack C), closed electrode + epoxy resin + microscope slide (Stack D).
Besides the top electrode and the encapsulation, all other layers were kept as shown in the Figure \ref{fig:layer-stack-scheme} \textbf{a} (Stack PE+$\mathrm{Alq_3}$), with a thickness of \SI{40}{nm} for the $\mathrm{Alq_3}$ layer.

Images of the fabricated devices and their schematic layer stack are depicted in Figure S9.
It should be noted that the closed electrode configuration increases the effective active area of the electrodes and hence the cross-section available for the charge transport in the organic layer, leading to an increase in the measured electrical current. As shown in Figure \ref{fig:thickness-variation-and-encapsulation} \textbf{b}, the sensor response strength scales with the active layer area, which is significantly enhanced in the case of a closed electrode. Furthermore, all devices exhibited a pronounced sensor response, demonstrating that encapsulation with epoxy resin and glass does not completely suppress sensor functionality. Nevertheless, encapsulation by an additional microscope slide leads to a reduced sensor response strength, likely due to the limited diffusion of hydrogen through the encapsulating glass. For all devices, a linear dependence of the sensor response on hydrogen concentration was observed. These results highlight that both device performance and long-term stability can be tuned upon the choice of the electrodes and the encapsulation strategy. Typically, for OLED devices, a fully closed electrode is used together with an epoxy resin and microscope slide encapsulation, limiting degradation in the atmospheric environment \cite{ghosh2005thin}.

\begin{figure*}
    \centering
    \includegraphics[width=1\linewidth]{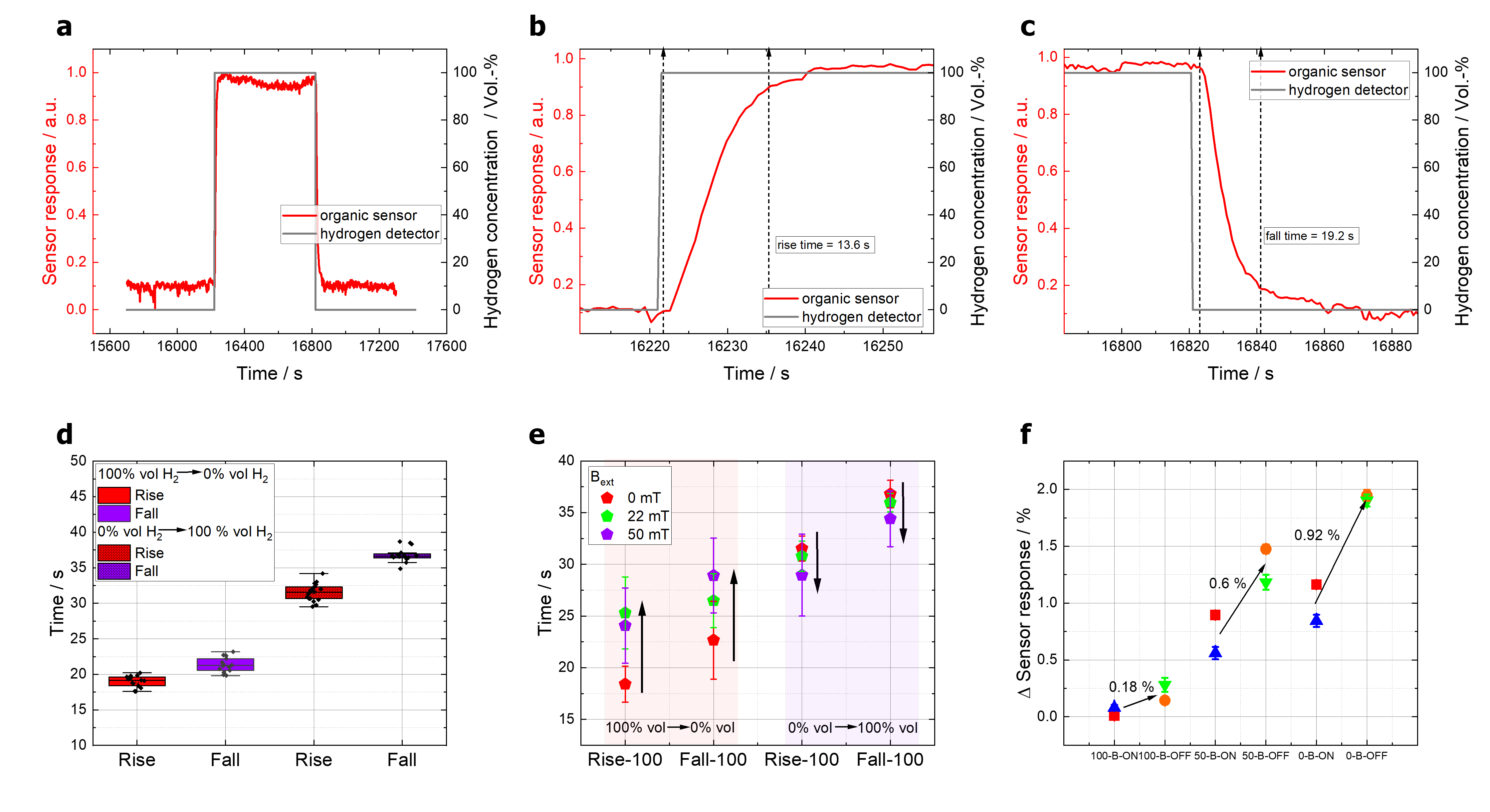}
    \caption{Evaluation of rise and fall times of the sensor response in panels \textbf{a–c}, with panel \textbf{a} showing a single cycle of the sensor response to hydrogen. In all three diagrams, the sensor response was inverted and normalized, such that the signal increases during the rise time and decreases during the fall time. However, the raw sensor signal itself decreases under hydrogen exposure due to the increase in resistance. In \textbf{b} the determination of the rise time is depicted, while the fall time evaluation is shown in \textbf{c}. Please note that the sensor response was reversed for this demonstration. The $t_{90}$, and $t_{10}$ times were determined for a relative change between $0\mathrm{\% \, vol}$ and $100\mathrm{\% \, vol}$ $\mathrm{H_2}$. In \textbf{d}, the rise and fall time for several measurements and devices are shown. In \textbf{e}, the dependence of the rise and fall times on the application of an external magnetic field is compared for increasing and decreasing hydrogen concentrations. \textbf{f} illustrates the effect of the magnetic field on the sensor signal determined at $100\mathrm{\% \, vol}$, $50\mathrm{\% \, vol}$, and $0\mathrm{\% \, vol}$ $\mathrm{H_2}$, respectively.}
    \label{fig:magneticfield dependence and rise-fall-time}
\end{figure*}

In Figure \ref{fig:magneticfield dependence and rise-fall-time} \textbf{a-c}, the procedure used for the determination of the rise and fall times is exemplarily shown. The $t_{90}$ (rise) time is defined as the time required for the sensor response value to change from $10 \, \%$ to $90 \, \%$ of the maximum value for a given hydrogen concentration, while the $t_{10}$ (fall) is defined \textit{vice versa} \cite{mandal2025robust}. Interestingly, the mean rise and fall times differ depending on whether the hydrogen concentration is increasing or decreasing. Thereby, decreasing the hydrogen concentration results in lower rise and fall time, $t_{90} = (18.4 \pm 1.7) \, \mathrm{s}$, and $t_{10} = (22.7 \pm 3.7) \, \mathrm{s}$ (cf. Figure \ref{fig:magneticfield dependence and rise-fall-time} \textbf{d}), respectively. For comparison, the $t_{90} = (31.5 \pm 1.2) s$, and $t_{10} = (36.8 \pm 1.3) \, \mathrm{s}$ were detected for increasing hydrogen concentration. In both cases, the difference between rise and fall time is very low, suggesting a highly reversible mechanism. 

Additionally, it was found that the application of an external magnetic field allows further tuning of the rise and fall times, as shown in Figure \ref{fig:magneticfield dependence and rise-fall-time} \textbf{e}. Organic materials are well known to exhibit changes in conductivity, resistivity, and electroluminescence under the application of an external magnetic field \cite{liu2009magnetic, morgenstern2025unlocking}. To probe these effects, we applied a static magnetic field to the sensor during hydrogen exposure. The magnetic field lifts the degeneracy of spin-dependent energy levels, thereby altering the spin-pair interconversion of polaron-pair (PP) species \cite{morgenstern2025unlocking, mondal2023degradation}. Thus, changes in conductivity and electroluminescence are observed.  A detailed analysis of magnetic field effect measurements (without the exposure to hydrogen) is provided in the SI (cf. Figure S10/S11 and the corresponding text). We observed that the rise and fall times could be effectively tuned by the magnetic field. While the magnetic field adversely affects the rise and fall times during decreasing hydrogen concentration, it enhances them when the hydrogen concentration is increasing. Comparing the influence of the magnetic field on the sensor response at different hydrogen concentrations reveals that the effect increases with decreasing hydrogen concentration (cf. Figure \ref{fig:magneticfield dependence and rise-fall-time} \textbf{f}). Magnetic fields obviously modify the sensor response. The hyperfine coupling of PP species dominates the spin interactions between singlet and triplet PP states in those $\mathrm{Alq_3}$ devices (details are provided in the SI, cf. Figure S10/S11). These observations indicate that hyperfine interactions contribute to the sensing mechanism, consistent with previous reports by Nguyen \textit{et al.} \cite{nguyen2007magnetoresistance}. Specifically, the presence of hydrogen atoms in bound organic compounds, and thus proton nuclear spins, appears to be a prerequisite for observing MFE. 

In conclusion, these experiments indicate that the gas-sensing behavior results from a combination of an interface effect between the active layer and the electrode, together with a bulk effect of the $\mathrm{Alq_3}$ layer. Nevertheless, as the effect saturates for increasing layer thickness of the active material, we suggest that the interface effect dominates the sensor signal. This interpretation is further supported by the observed improvement in rise and fall times in the presence of $\mathrm{Alq_3}$ compared to pure $\mathrm{PEDOT:PSS/MoO_3}$. 
\label{Sensor Principle}
\begin{figure*}
    \centering
    \includegraphics[width=1\linewidth]{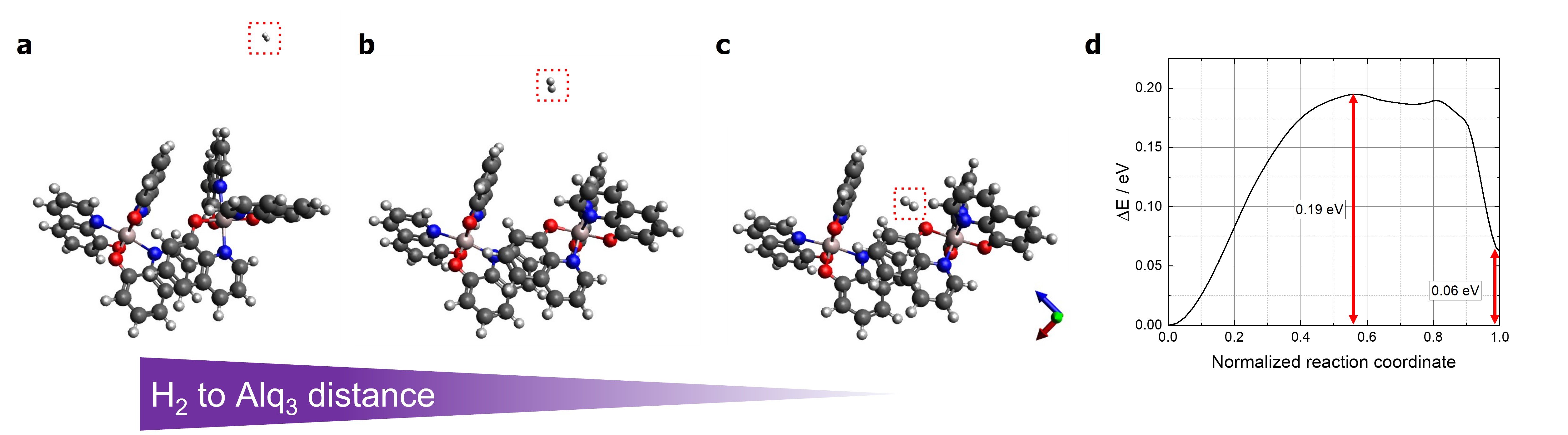}
    \caption{Simulation of the impact of molecular hydrogen on $\mathrm{Alq_3}$ geometry using DFT calculations with varying $\mathrm{H_2}$–$\mathrm{Alq_3}$ separation. \textbf{a-c} Decreasing distance between $\mathrm{H_2}$ and $\mathrm{Alq_3}$, leading to pronounced molecular deformation. Such deformation is expected to affect charge transport and increase resistivity. \textbf{d} Energy profile of hydrogen incorporation, referenced to the energy of $\mathrm{H_2}$ far from the $\mathrm{Alq_3}$ being chosen as the zero point. $\Delta E$ denotes the energy difference relative to the initial state. The reaction barrier is broad and approximately \SI{0.19}{eV} high, and the final structure is about \SI{0.06}{eV} higher in energy than the initial structure.}
    \label{fig:DFT}
\end{figure*}

To date, the underlying physical mechanism of the sensor remains unconfirmed. We propose that the response arises from a combination of bulk and interface effects: the bulk contribution results from the diffusion of hydrogen molecules through the organic layer, while the interface contribution originates from the charge-injection barrier between the organic layer and the bottom electrode. In the following, the proposed mechanisms will be explained in detail. 
As previously mentioned, catalytic materials are required when molecular hydrogen is dissociated into atomic hydrogen \cite{mandal2025robust}. In the study from Mandal \textit{et al.}, a Pt electrode was used for this purpose. As atomic hydrogen is much more reactive, physisorption or even chemisorption processes are reasonable. Similar processes were also found for metal oxide-based sensors \cite{ji2019gas}. For those sensor principles, however, oxygen is required for the chemical reaction and thus to regenerate the sensor. As we performed all experiments (besides cross-sensitivity) in an inert atmosphere, we can exclude this chemical reaction.

To prove the integrity of the molecules before and after exposure to hydrogen gas for two hours, photoluminescence (PL) and atomic-force microscopy (AFM) measurements were performed on thin films containing an evaporated \SI{40}{nm} thick $\mathrm{Alq_3}$ layer without any encapsulation. The results are shown in the SI in Figure S12 \textbf{a-c}. The morphology did not change when comparing the thin film before and after hydrogen exposure. Assuming a chemical reaction of the molecule and the hydrogen atoms, one would expect that this is not easily reversible. Furthermore, the PL spectra before and after hydrogen exposure are identical, an indication of the reversibility of the eventual changes in the molecular electronic structure occurring during the hydrogen exposure.

Additionally, the electroluminescence of the PE+$\mathrm{Alq_3}$ sensor was monitored during exposure to hydrogen, and no changes of the EL spectra were observed (cf. Figure S12 \textbf{d}). The "bulk" mechanisms discussed above only explain the increase in the sensor resistance, \textit{i.e.} the negative sensor response. The decreasing resistance for the Stack PE must hence be dominated by the contribution of the ITO/$\mathrm{PEDOT:PSS/MoO_3}$ interface rather than by the distortion of the molecular geometry.\newline
Conclusively, the impact of the bulk effect in $\mathrm{Alq_3}$ seems to be rather small compared to the interface contribution.

A plausible explanation for the opposite sign of the resistance response in $\mathrm{Alq_3}$ and $\mathrm{PEDOT:PSS:MoO_3}$ under hydrogen involves different interface-dipole interactions. In Stack PE devices, hydrogen modifies the ITO surface in a way that it lowers the hole-injection barrier, as it is pinned to the work function of ITO. Thus, increasing the number of holes at the electrode side decreases the resistance. In the full stack, $\mathrm{Alq_3}$ produces a net increase in the injection/transport barrier, leading to fewer mobile carriers or additional energy barriers, which results in a decrease in resistance. One has to note that this scenario requires the dissociation of molecular hydrogen. As oxygen is not available, a reaction is assumed that only results in additional charge carriers.

As previously mentioned, according to the observed sensor response dependence on the thickness of the $\mathrm{Alq_3}$ layer, a bulk contribution cannot be ruled out. Accordingly, quantum chemical calculations were conducted. The proposed mechanism responsible for the hydrogen sensing in $\mathrm{Alq_3}$ is depicted in Figure \ref{fig:DFT}. The distance between two molecules and the molecular hydrogen was consistently varied (cf. Figure \ref{fig:DFT} \textbf{a} to \textbf{c}, decrease in distance). As can be seen, the molecular interaction changes when $\mathrm{H_2}$ diffuses between two molecules. This deformation is expected to cause changes in electrical conductivity, lowering overall device performance. The molecular deformation is easily reversible. This explains the low difference between the rise and fall time of our sensors, compared to previously reported values, where the rise time was reported to be approximately $ 35-90 \ \%$ \cite{mandal2025robust, girma2023room, xing2022three} lower compared to the fall time. For our molecules, the two values differ by only $(15.5 - 20.9) \, \%$.

Time-dependent density functional theory (TD-DFT) calculations also revealed that the $\mathrm{Alq_3}$ absorption spectrum remains unchanged upon exposure to hydrogen. The first three calculated absorption bands for the initial geometry are \SI{3.00}{eV}, \SI{3.09}{eV}, and \SI{3.16}{eV}, while those for the final geometry are \SI{3.02}{eV}, \SI{3.05}{eV}, and \SI{3.12}{eV}. These differences of \SI{0.02}{eV}, \SI{0.04}{eV}, and \SI{0.04}{eV} are far below the range that TD-DFT can accurately differentiate and can therefore be considered unchanged after hydrogen incorporation. Additionally, the activation barrier was determined using the nudged elastic band method (NEB) \cite{ORCA6.1Manual}. The activation barrier refers to the energy required to deform the molecule.
According to the Eyring equation of transition state theory, the rate of a reaction in relation to the energy barrier is described as follows:
\begin{equation}
k = \alpha \frac{k_B T}{h} , e^{-\frac{\Delta G^{\ddagger}}{k_B T}},
\end{equation}
where $h$ is Planck constant, $k_\mathrm{B}$ the Boltzmann constant, $\alpha$ a transmission coefficient representing the fraction of molecules in the transition state that proceed to the final state, and $\Delta G^{\ddagger}$ the Gibbs free energy difference between the initial and the transition states, and thus corresponding to the activation barrier. The symbol $^{\ddagger}$ denotes the transition state. Figure \ref{fig:DFT} \textbf{d} shows the resulting energy profile of the NEB calculations. It is important to note that the shown energy difference corresponds only to the electronic energy, instead of the Gibbs energy. To get the Gibbs free energies of the two states, the vibrational frequencies using the structures from the NEB calculation need to be determined. This gives the vibrational, thermal, and energy terms of the total energy. This Gibbs free energy difference is slightly larger than the purely electronic energy difference from the NEB method at $\approx \SI{0.28}{eV}$ ($\SI{6.4}{kcal/mol}$). With the assumption that the transition coefficient is one, we get a rate constant of $ \approx 1.3 \cdot10^8 \, \mathrm{s^{-1}}$, corresponding to a theoretical half-life for the reaction of \SI{5.4}{ns}. According to DFT calculations, the reaction is expected to proceed almost instantaneously, in agreement with the minimal difference between rise and fall times and their overall low magnitudes.

\subsection{Operational Robustness} 

\begin{figure*}
    \centering
    \includegraphics[width=0.7\linewidth]{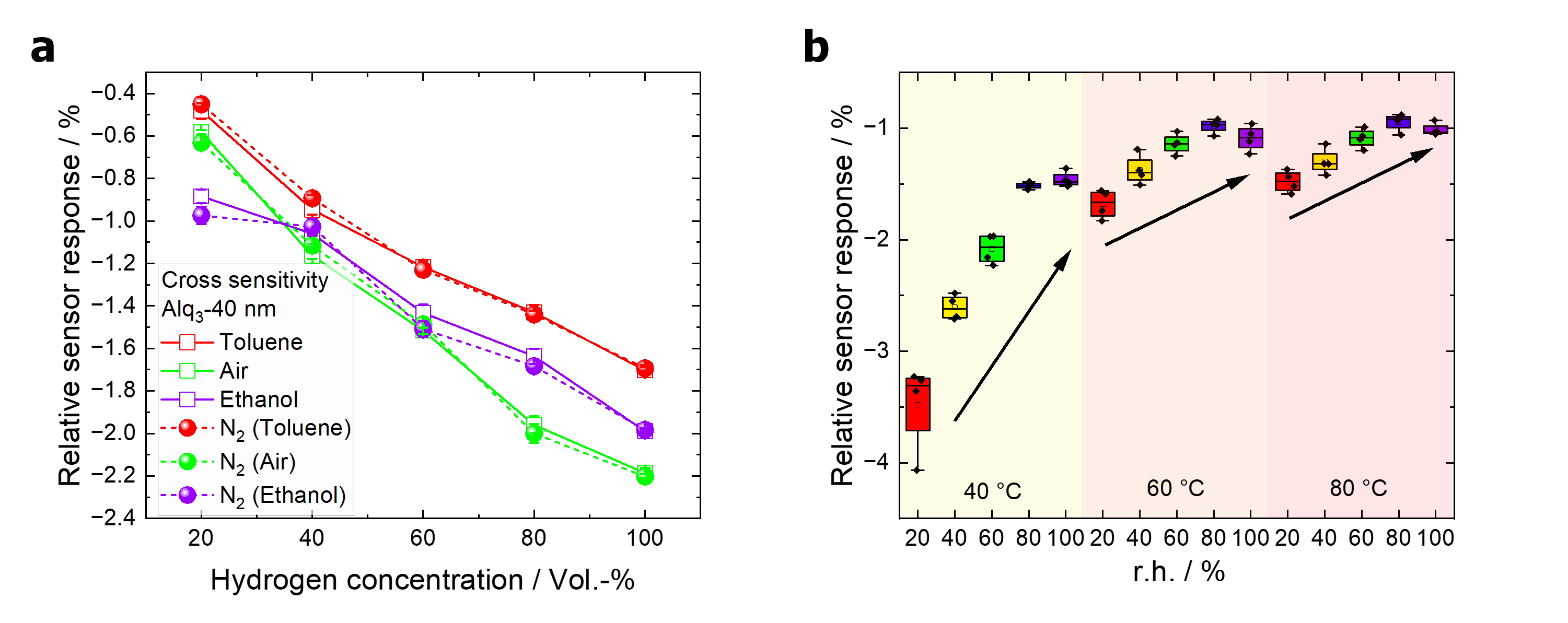}
    \caption{ \textbf{a} Cross-sensitivity to toluene, ethanol, and dry air as a function of hydrogen concentration. \textbf{b} Temperature and humidity dependence at a hydrogen concentration of $100 \, \mathrm{vol}\%$}. 
    \label{fig:encapsulation-cross-sencistivity-temp}
\end{figure*}

The sensors cross-sensitivity to other gases was evaluated for the devices with a structured electrode together with an epoxy encapsulation using dry air, ethanol, and toluene. The corresponding results are presented in Figure \ref{fig:encapsulation-cross-sencistivity-temp} \textbf{a}. The hydrogen concentration was varied from $100\mathrm{\% \, vol} \ \mathrm{H_2}$ to $0\mathrm{\% \, vol} \ \mathrm{H_2}$ in steps of $20\mathrm{\% \, vol}$, while alternating the carrier gas between nitrogen and the test gas. No significant increase or decrease in device current was observed during the carrier-gas exchange for air and ethanol, indicating negligible cross-sensitivity toward these gases. In contrast, the curve obtained for toluene shows a slight decrease in the sensor signal. Notably, the signal does not recover after switching back to nitrogen, suggesting that toluene exposure leads to a small but persistent decrease in relative sensor response that should be taken into account. The raw data can be found in the SI in Figure S13. 

Especially for the application in a fuel cell system, the sensor has to exhibit a reliable signal under fuel cell like conditions, meaning an operating temperature up to $80 ^\circ$C and a humidity of $100 \ \% \ \mathrm{r.h.}$ \cite{li2013high, xiao2016ion, zang2014specific}. The sensor response was measured for three different temperatures ranging from $40^\circ$C to $80^\circ$C. For each temperature, the humidity was increased from $20 \ \% \ \mathrm{r.h.}$ to $100 \, \% \ \mathrm{r.h.}$, respectively. The dependence of the sensor signal on humidity and temperature is shown in Figure \ref{fig:encapsulation-cross-sencistivity-temp} \textbf{b}.
Humidity has a clear influence, yet the effect is still distinctly observable at $80 \, ^\circ\text{C}$. As expected for organic molecules, high humidity limits charge carrier transport \cite{aziz1998humidity}, thus limiting the sensor response. Interestingly, slightly increasing the temperature from RT to $40 \, ^\circ\text{C}$ improves the sensor response, which might be explained by thermal activation of trapped charges \cite{kohler2015electronic}. 
By further increasing the temperature up to $80 \, ^\circ\text{C}$ and $100 \, \% \ \mathrm{r.h.}$, the sensor signal is decreased to $\approx 1 \, \%$, but the sensor remains functional. It has previously been found that the charge carrier mobility in $\mathrm{Alq_3}$ decreases at elevated temperatures. This reduction in mobility may limit the sensor response at higher temperatures \cite{park2006temperature}. 

The protection provided by the resin against humidity is less efficient when increasing the temperature for a given humidity, on one hand, and when increasing the humidity at a given temperature, on the other hand.
For better protection against humidity, various encapsulation strategies can be employed. For further studies, the use of temperature-stable epoxy resins appears particularly promising to ensure long-term environmental stability \cite{kandola2010studies, ehlers1960correlation}.

\FloatBarrier
\section{Conclusion and Outlook}
In conclusion, we demonstrated a novel, robust, and reproducible hydrogen sensor principle based on organic molecules that exhibit high sensitivity and linear sensor response to hydrogen concentration in a large concentration range, ranging from $0\% \, \mathrm{vol}$ to $100\% \, \mathrm{vol}$. The devices employ a vertical stack geometry, though the concept is not limited to this architecture, and crucially, the sensor response does not rely on radiative recombination. While the underlying sensing mechanism is still under debate, our analysis suggests a combination of bulk and interfacial effects, within the active layer and at the interface towards the bottom electrode. Notably, the sensor operates without requiring oxygen for regeneration, making it particularly suitable for safety-critical applications and closed systems, including enclosed fuel cell environments. The response time in ambient atmosphere was found to be low, with typical $t_{90} \approx 18$ s, and $t_{10} \approx 23$ s. A significant cross-sensitivity towards ethanol and dry air was not observed. However, toluene slightly quenches the strength of the sensor response. Temperature and humidity studies indicate that encapsulation improvements are necessary to ensure stable operation under high-humidity conditions, a critical consideration for practical deployment in fuel cell systems. A robust sensor response was found up to a temperature of $80 ^\circ$C and $100 \, \%$ r.h.
Overall, this work establishes a new direction for hydrogen sensing, offering a robust, oxygen-independent, and scalable platform that has the potential to enhance safety, monitoring, and environmental sustainability across hydrogen-based technologies.

\section{Methods}
\subsection{Sensor preparation}
The $\mathrm{Alq_3}$ sensor devices were fabricated with the following structure (\textit{cf.} Figure~\ref{fig:layer-stack-scheme}\textbf{a}): ITO (\SI{115}{nm}, purchased from Lumtec; pre-patterned cathode with six pixels) / PEDOT:PSS + $\mathrm{MoO_3}$ (\SI{35}{nm}) / $\mathrm{Alq_3}$ (\SI{50}{nm}) / LiF (\SI{0.8}{nm}) / Al (\SI{110}{nm}).

For the 4CzIPN devices, the following layer stack was used: ITO / PEDOT:PSS + $\mathrm{MoO_3}$ (\SI{35}{nm}) / 4CzIPN (\SI{30}{nm}) / TPBi (\SI{20}{nm}) / LiF (\SI{0.8}{nm}) / Al (\SI{110}{nm}).

PEDOT:PSS was purchased from Heraeus (Clevios™ PEDOT:PSS). 2,2$'$\,2$''$-(1,3,5-benzenetriyl)-tris(1-phenyl-1H-benzimidazole) (TPBi) was purchased from Ossila. $\mathrm{Alq_3}$, $\mathrm{MoO_3}$ powder, and LiF were purchased from Sigma-Aldrich. 1,2,3,5-tetrakis(carbazol-9-yl)-4,6-dicyanobenzene (4CzIPN) was purchased from Ossila.

$\mathrm{MoO_3}$ powder was dispersed in an ammonium hydroxide solution (\SI{0.25}{g.mL^{-1}} $\mathrm{NH_4OH}$) and added at a mass ratio of 1:0.02.
% The $\mathrm{Alq_3}$ sensor devices were fabricated with the following structure (\textit{cf.} Figure \ref{fig:layer-stack-scheme} \textbf{a}): ITO (\SI{115}{nm}), purchased by Lumtec, pre-patterned cathode (six pixels)/ PEDOT:PSS + $\mathrm{MoO_3}$ (\SI{35}{nm})/ $\mathrm{Alq_3}$ (\SI{50}{nm})/ LiF (\SI{0.8}{nm})/ Al (\SI{110}{nm}). For the 4CzIPN devices, the following layer stack was used: ITO/ PEDOT:PSS + $\mathrm{MoO_3}$ (\SI{35}{nm})/ 4CzIPN (\SI{30}{nm})/ TPBi (\SI{20}{nm})/ LiF (\SI{0.8}{nm})/ Al (\SI{110}{nm}). PEDOT:PSS was purchased from Heraeus (Clevios™—PEDOT:PSS). 2,2′,2″-(1,3,5-Benzinetriyl)-tris(1-phenyl-1-H-benzimidazole) (TPBi) was purchased from Ossila. $\mathrm{Alq_3}$, $\mathrm{MoO_3}$ powder, and LiF were purchased from Sigma-Aldrich. 1,2,3,5-Tetrakis(carbazol-9-yl)-4,6-dicyanobenzene (4CzIPN) was purchased from Ossila. $\mathrm{MoO_3}$ powder was dispersed in an ammonium hydroxide solution (0.25 g/ml $\mathrm{NH_3OH}$) was added at a mass ratio of 1:0.02. 

The thicknesses of the layers were measured using profilometry, with an associated thickness variation of \(8\%\). Before the PEDOT:PSS + $\mathrm{MoO_3}$ was deposited, the samples were cleaned with alkaline detergent + deionized water and isopropanol in an ultrasonic bath for $30$ min, respectively. Afterward, the samples were dried with nitrogen. The surface was further activated by a UV-ozone treatment with a custom-built setup (after \cite{weber2023cost}) for $35$ min. PEDOT:PSS + $\mathrm{MoO_3}$ was spin-coated at \SI{5000}{rpm} (acceleration $500 \ \mathrm{rpm/s}$) for $30$ s and post-annealed at $130$ $^\circ$C for 30 min. $\mathrm{Alq_3}$ was evaporated at constant rates of $1 \, \mathrm{\r{A}}$/s. 4CzIPN, and TPBi were also evaporated at constant rates between $1-1.5 \,\mathrm{\r{A}}$/s, respectively.  LiF and Al were evaporated at constant rates of 0.1 $\mathrm{\r{A}}$/s and 5 $\mathrm{\r{A}}$/s, respectively. The deposition via thermal evaporation in the vacuum chamber was performed at a pressure of $10^{-6} \, \mathrm{mbar}$. The shadow masks were fabricated from \SI{300}{\micro m}, double-sided polished, \SI{150}{mm} silicon wafers with a $\{100\}$ orientation. A \SI{1000}{nm} thermal oxide layer was first grown on the entire wafer surface. A photoresist was then deposited on the rear side of the wafer to pattern the shadow mask structure. The silicon dioxide ($\mathrm{SiO_2}$) was subsequently wet-etched, followed by a \SI{50}{\micro m} deep anisotropic silicon etching using the Bosch process. The photoresist was removed afterward using an oxygen plasma.
Next, plasma-enhanced chemical vapor deposition (PE-CVD) $\mathrm{Si_3N_4}$ was deposited on the front and rear side of the wafer to serve as a mask for potassium hydroxide (KOH) etching. Photolithography was performed on the front side to open the silicon nitride ($\mathrm{Si_3N_4}$) mask via reactive ion etching (RIE), while the $\mathrm{SiO_2}$ layer was removed by wet etching. This was done in preparation for subsequent KOH etching down to \SI{270}{\micro\meter} to open the shadow mask structures. Finally, the remaining $\mathrm{Si_3N_4}$ and $\mathrm{SiO_2}$ layers were removed by RIE and wet etching, respectively. The devices were encapsulated with epoxy resin via blade coating or drop coating (specified for two sets of samples mentioned in the text). For two investigations throughout this work, an additional microscope slide was placed on top of the epoxy resin. In all cases, the samples were post-treated by UV illumination for 15 min to cure the resin.
\subsection{Characterization}
The hydrogen sensors were electrically characterized by Keithley devices of the type Keithley2636A and Keithley2636B. To detect the outcoupled light from the sensing device, an additional photodiode from Hamamatsu S11 was used, where the photocurrent was monitored throughout the measurement. The electrical measurements were controlled by custom Python programs. The exposure to hydrogen was achieved by a HovaCal® N 1464-SP calibration gas generator. A constant flow rate of \SI{1100}{ml/min} was chosen for all experiments. It offers a mixture of a carrier gas (here, nitrogen) and hydrogen. For cross-sensitivity measurements, the toluol and ethanol were evaporated and mixed with hydrogen. Dry air was tested by exchanging the carrier gas. Temperature and humidity investigations were performed in the climate chamber DISCOVERY DM1600 C(T) from ACS. The temperature varied between room temperature and $\mathrm{80 ^\circ C}$, while the humidity varied between $\mathrm{0\% \, vol}$, and $\mathrm{100\% \, vol}$. Magnetic field effect measurements were performed using a custom-built measurement setup. All details can be found elsewhere \cite{morgenstern2025unlocking}. AFM images were taken by an AJST-NT SPM $\mathrm{SmartSPM^{TM}}$-100 AFM. The microscopy images were taken using an optical microscope, namely Nikon IC inspection Microscope ECLIPSE L200 from Nikon Metrology GmbH (Düsseldorf, Germany) with 1×, 5x, and 20× objectives. Profilometry measurements were conducted by a stylus profilometry VEECO Dektak.
\subsection{DFT methodology}
All DFT calculations were performed using the ORCA software package version 6.1 \cite{neese2023shark, neese2025software}. The conformers of two $\mathrm{Alq_3}$ molecules were calculated with the GOAT algorithm implemented as part of the ORCA package \cite{de2025goat} at the GFN2-XTB level \cite{bannwarth2019gfn2, bannwarth2021extended}. The lowest conformers were optimized with the wB97X-3c composite method \cite{wittmann2024extension}, which is optimized for the calculation of reaction barriers and thermochemistry. The calculations with these hybrid functionals were sped up using the resolution of identity \cite{neese2002efficient, neese2003improvement} and chain of spheres \cite{neese2009efficient, helmich2021improved} approximations. During this and the following equations, the dielectric environment of a molecular film was approximated with the CPCM model \cite{garcia2020effect}. The lowest energy conformer was chosen for the NEB calculation. The reactant and product configurations were generated by respectively placing a hydrogen molecule far away from the $\mathrm{Alq_3}$ molecules and between the $\mathrm{Alq_3}$  molecules. Afterwards, both structures were reoptimized. NEB calculations \cite{asgeirsson2021nudged, schmerwitz2023improved} were performed using eight intermediate images. The resulting transition state structure was extracted, and the reactants vibrational properties were calculated \cite{bykov2015efficient, garcia2019efficient}. Finally, TD-DFT calculations using the PBE0 global hybrid functional \cite{adamo1999toward} were performed for both reactant and product structures after reoptimization with this functional. For this, the def2-TZVP triple-zeta basis set \cite{weigend2005balanced} and the D4 dispersion correction \cite{caldeweyher2017extension, caldeweyher2019generally, caldeweyher2020extension, wittmann2024extension} were employed.

\subsection{Acknowledgments}
The authors would like to thank Thomas Seyller for providing access to the profilometer.
\newline
A.M., L.V., T.v.U., and G.S. would like to thank SAB for funding this research under the project number 100649391 (ReSIDA-H2). The authors gratefully acknowledge funding by the Deutsche Forschungsgemeinschaft (DFG, German Research Foundation) through DFG-TRR 386-B05 (514664767). F.O. would like to thank the DFG for financial support through the Cluster of Excellence e-conversion (Grant No. EXC 2089 - 390776260).

% \FloatBarrier
% \newpage
% \bibliographystyle{MSP}
% \bibliography{sn-bibliography}% common bib file

% \begin{document}
% \FloatBarrier
% \title{
% \centering
% Supplementary Information\\
% Organic Hydrogen Sensors for Potential Use in Safety-Critical Environments
% }

\begin{figure}[H]
    \centering
    \includegraphics[width=0.7\linewidth]{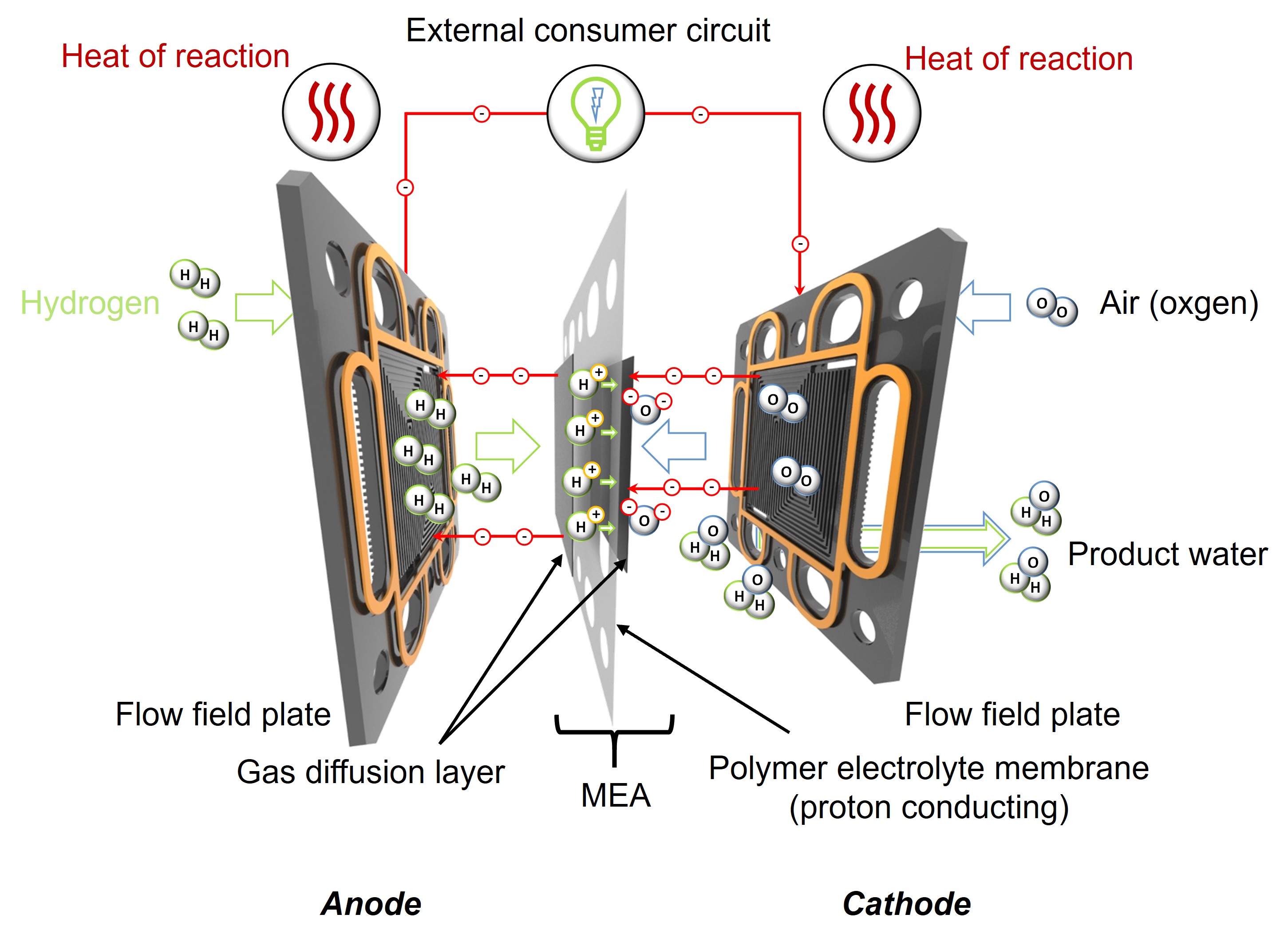}
    \caption{Schematic of a single PEM fuel cell showing its typical structure and operating principle. Hydrogen is supplied to the anode and oxygen to the cathode. The membrane electrode assembly (MEA) between the flow field plates facilitates proton conduction, while the electrochemical reaction produces electrical energy and heat, and water as a by-product. }
    \label{fig:PEMFC}
\end{figure}

\newpage
\maketitle
\supplementarysection
To better understand the specific requirements for hydrogen sensors in fuel cell systems, it is useful to first consider the fundamental operating principle of a proton exchange membrane fuel cell (PEMFC), which is widely used in both mobile and stationary applications \cite{en17205158, daud2017pem}. Figure \ref{fig:PEMFC} shows a schematic representation of a single PEMFC and its typical operating principle \cite{keller2023beitrag}. A PEMFC converts the chemical energy of hydrogen and oxygen directly into electrical energy. Oxygen is supplied to the cathode of the fuel cell, while hydrogen is fed to the anode, where it dissociates into protons and electrons. While the protons migrate through the polymer electrolyte membrane, the electrons travel through an external circuit, generating an electric current. At the cathode, the protons and electrons recombine with oxygen to form water. This process continues as long as hydrogen is supplied, and produces only water and heat as byproducts \cite{larminie2003fuel}.

Considering the evolution of hydrogen concentration and the corresponding cell voltage in a typical PEMFC system, over time, nitrogen and small amounts of other inert gases (\textit{e.g.} argon) accumulate in the anode chamber, primarily via diffusion from the cathode, leading to a dilution of the hydrogen stream. This results in a decrease of the hydrogen partial pressure, which causes a drop in cell voltage and thus a loss in system efficiency.

To mitigate this effect, the purge valve is periodically opened to expel the accumulated inert gases and restore hydrogen purity, thereby recovering cell voltage. However, this process also purges unconsumed hydrogen, which negatively impacts overall fuel cell efficiency.

Achieving an optimal balance between voltage stability and hydrogen consumption requires precise control of purge intervals. This, in turn, demands sensors capable of reliably monitoring the local hydrogen concentration under dynamic, in operando fuel cell conditions. The data provided by such sensors is essential for implementing intelligent purge strategies that minimize both performance losses (\textit{e.g.} fuel cell starvation) and hydrogen waste. Accordingly, sensor systems that require oxygen for regeneration are unusable for integration in fuel cell systems.

\FloatBarrier
\begin{figure}[H]
    \centering
    \includegraphics[width=0.8\linewidth]{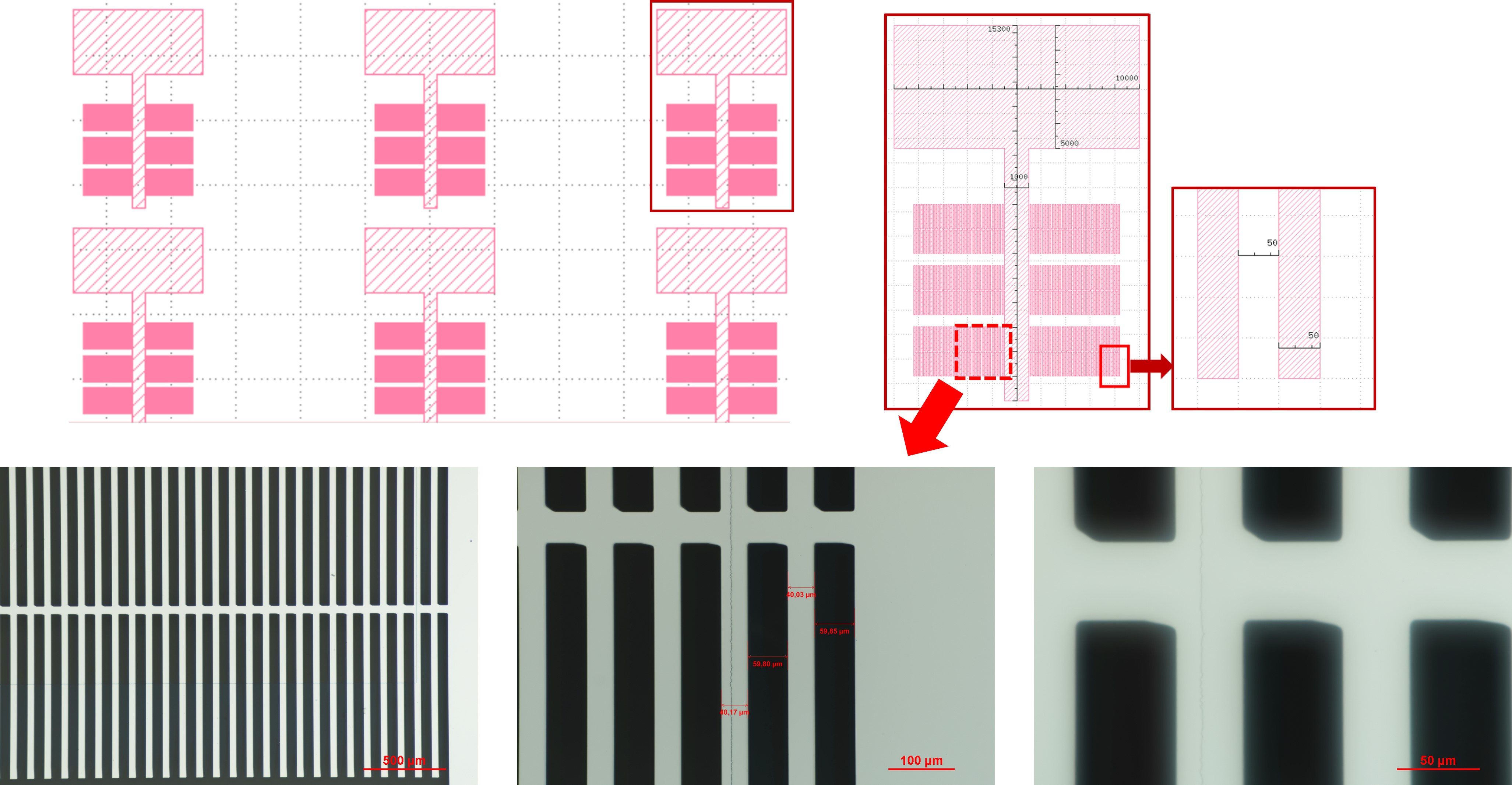}
    \caption{Mask layout for evaporating the top electrode aluminum and the underlying layer of lithium fluoride. The bars have a spatial distance of \SI{50}{\micro m} and the mask is made of silicon with a thickness of \SI{50}{\micro m} too. In addition to the mask layout, optical microscopy images of the evaporated structures are provided. }
    \label{fig:mask-layout}
\end{figure}

\begin{figure}[H]
    \centering
    \includegraphics[width=0.7\linewidth]{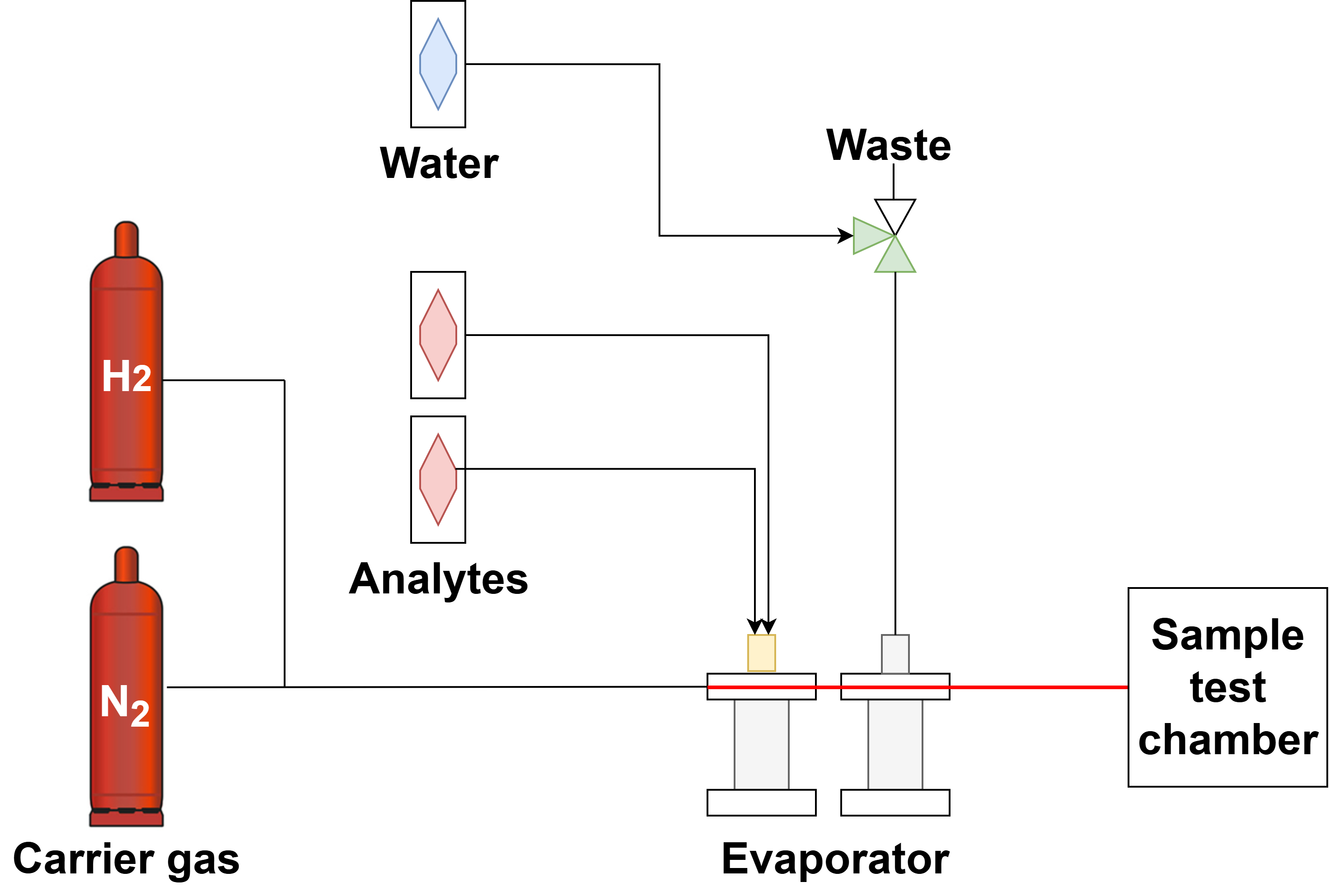}
    \caption{The gas mixing system is schematically shown here. It contains several connections for the gas inlet. Here we only used hydrogen and nitrogen. Further, analytes are available, which were used to evaporate toluene and ethanol. Another analyte is available for the evaporation of water, enabling humidity-dependent measurements. The gas mixing system was connected to the measurement chamber shown in Figure 1 \textbf{b}.}
    \label{fig:gas mixing system scheme}
\end{figure}

\begin{figure}[H]
    \centering
    \includegraphics[width=1\linewidth]{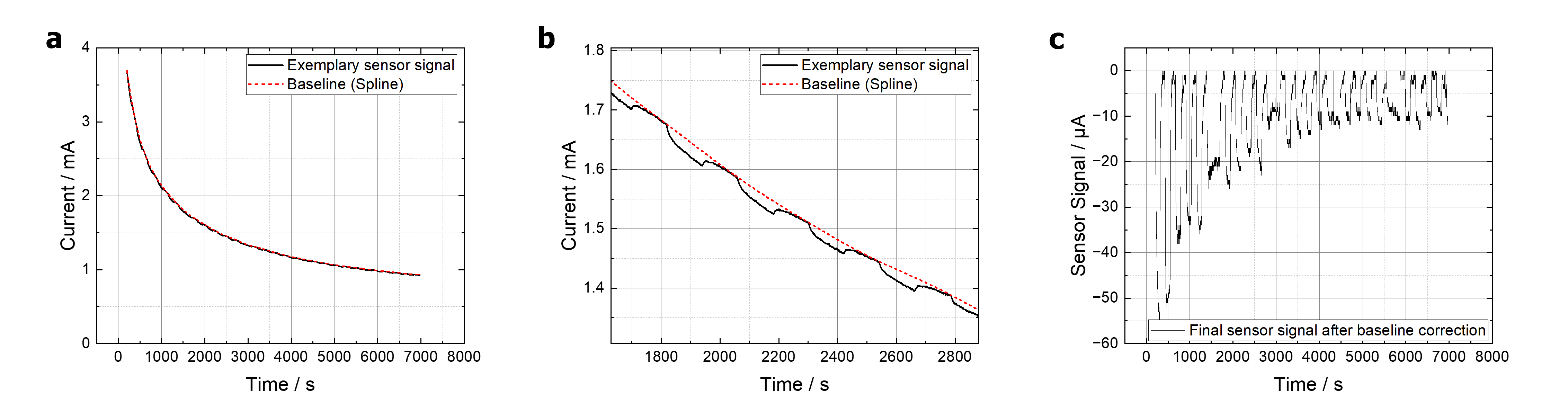}
    \caption{\textbf{a} shows the raw data along with the baseline derived from the spline function IRSQR, with a detailed view provided in the inset \textbf{b}. As seen, the baseline accurately captures the signal degradation in $I(\mathrm{t})$ over time without reflecting the hydrogen sensing response. In \textbf{c}, the baseline-corrected sensor signal is displayed, calculated as $I(\mathrm{gas\mbox{-}conc.)} = I\mathrm{(t + gas\mbox{-}conc.)} - I(\mathrm{t})$.} 
    \label{fig:baseline correction}
\end{figure}

\FloatBarrier
Figure \ref{fig:baseline correction} exemplarily shows a baseline correction using a spline function from the pybaselins library \cite{rapp_pybaselines}, namely IRSQR. The function is defined as follows. 
\begin{equation}
\min_{f} \sum_{i=1}^{n} \rho_\tau(y_i - f(x_i)) + \lambda \int \left( f''(x) \right)^2 dx
\end{equation}
The baseline is set to the upper quantile (here, $\tau $ = 0.9999), while the smoothing parameter $\lambda$ was chosen to be $1 \cdot 10^{-5}$. The other parameters were kept at the default values. the number of knots was set to 100, the maximum number of iterations to 100, and the tolerance for convergence determination to $1 \cdot 10^{-6}$. The order of the differential matrix was chosen to be 3. A full explanation can be found in the GitHub documentation \cite{rapp_pybaselines}.

\FloatBarrier
\begin{figure}[H]
    \centering
    \includegraphics[width=0.8\linewidth]{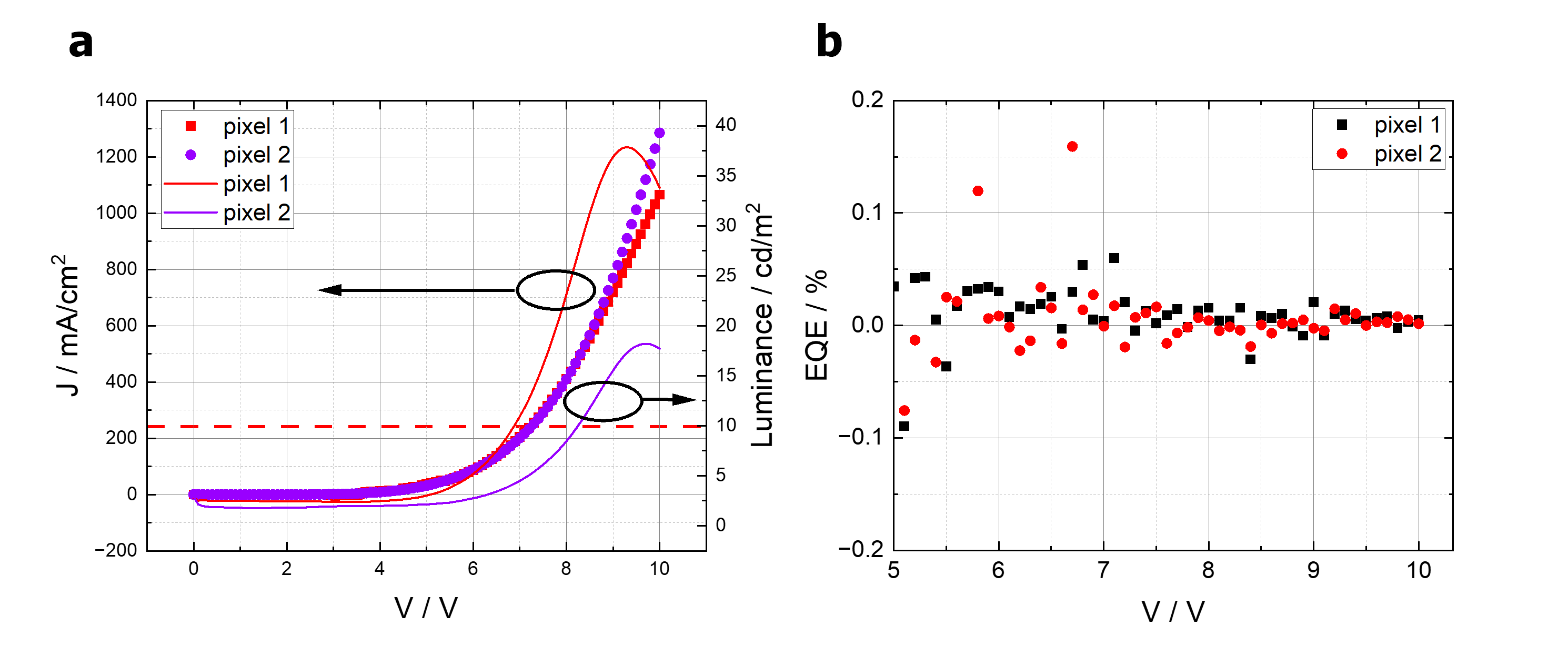}
    \caption{The optoelectronic characterization for the device based on $\mathrm{Alq_3}$ is provided, showing in \textbf{a} the current density and the corresponding luminance in a range of the applied bias voltage from 0 to 10 V. The red dotted line shows the marks the turn-on voltage (at $10 \ \mathrm{cd/m^2}$). In \textbf{b} the external quantum efficiency (EQE) is shown. Due to the simple layer stack on one hand and the structured top electrode on the other hand, the external quantum efficiency is quite low. Nevertheless, the light output does not play any role in the sensing properties, as explained in the manuscript.}
    \label{fig:Optoelectronic-characterization}
\end{figure}

\begin{figure}[H]
    \centering
    \includegraphics[width=0.6\linewidth]{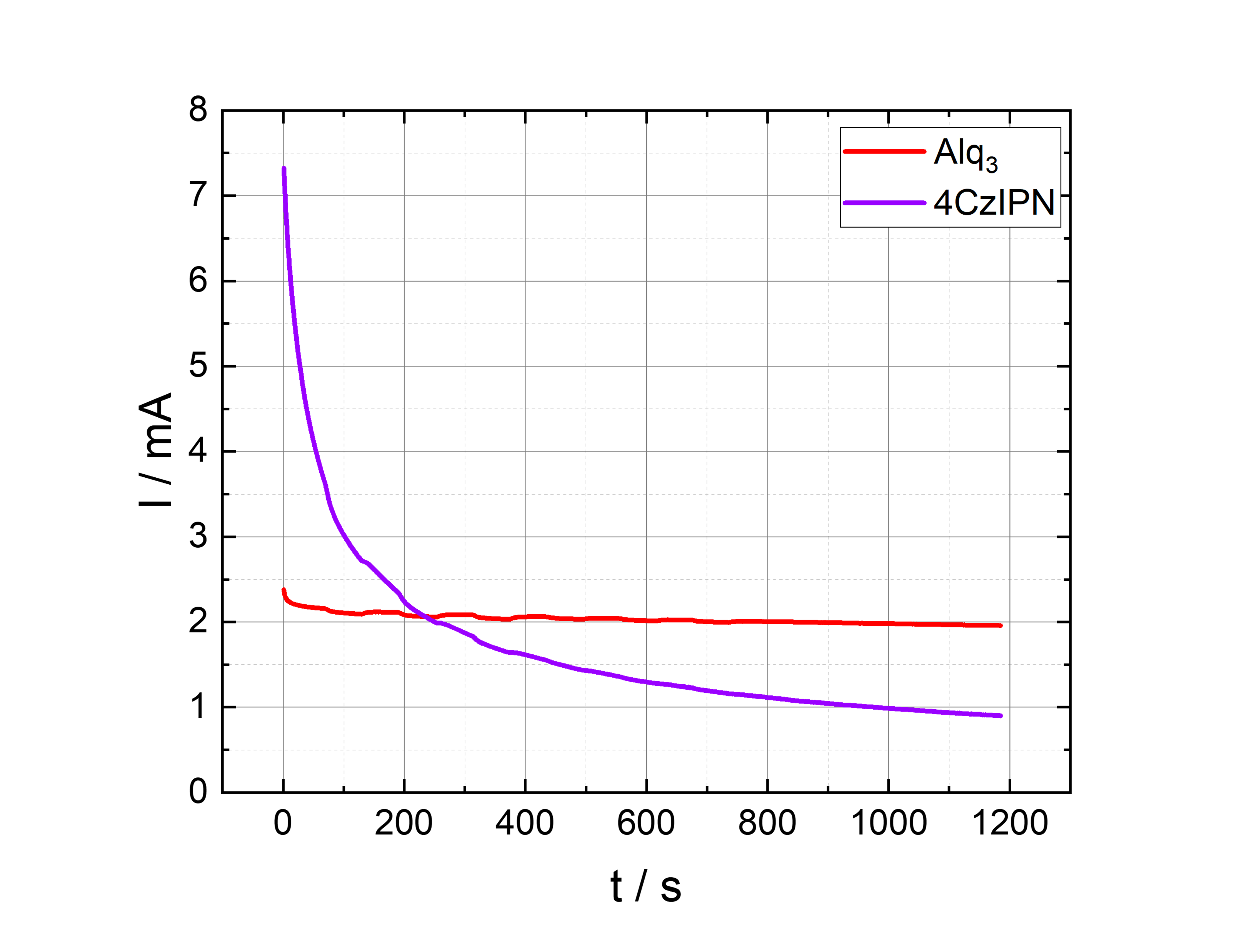}
    \caption{Comparison of the absolute current for a device containing $\mathrm{Alq_3}$ as active material and one containing 4CzIPN. The exact layer stack can be found in the manuscript in Figure 1  \textbf{a} and 2 \textbf{e}, respectively.}
    \label{fig:Optoelectronic-characterization}
\end{figure}

\begin{figure} [H]
    \centering
    \includegraphics[width=0.8\linewidth]{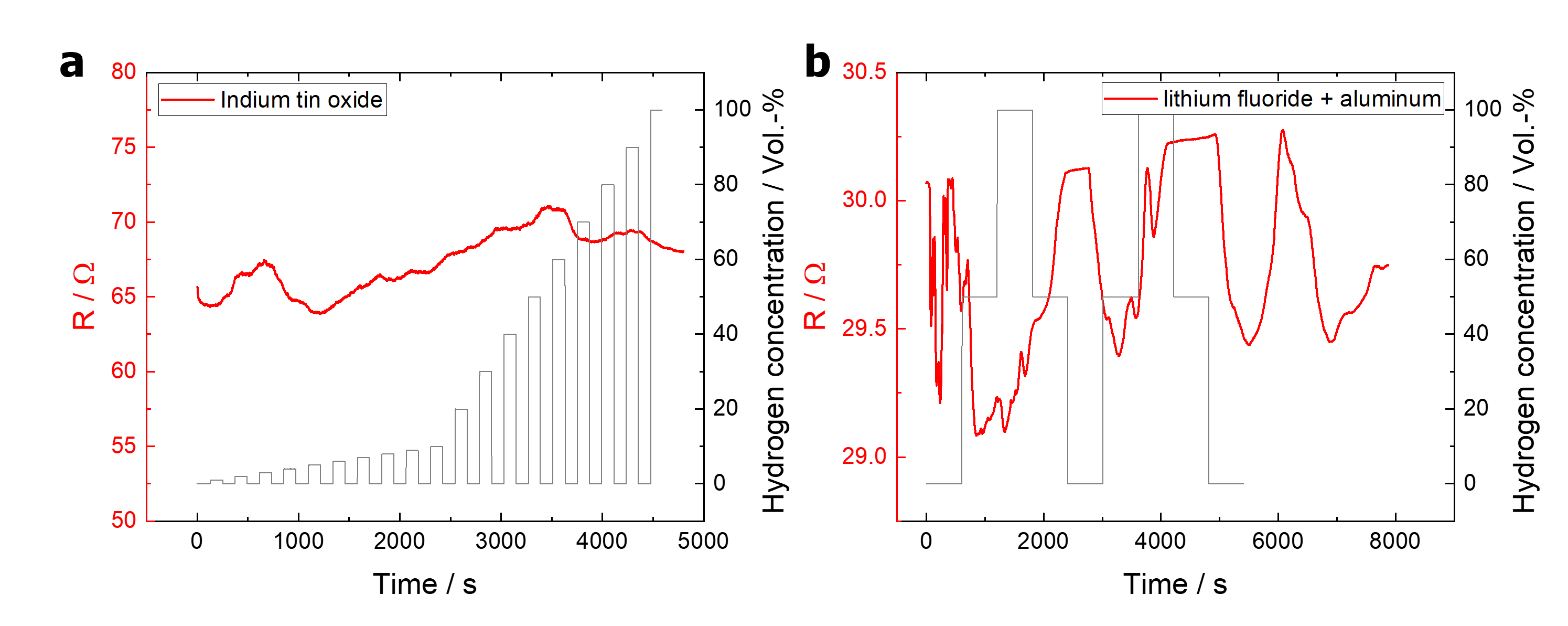}
    \caption{Resistance of \textbf{a} ITO, and \textbf{b} of lithium fluoride together with aluminum under hydrogen exposure. In both cases, a difference in resistance upon the hydrogen exposure was not monitored, excluding a bulk influence.}
    \label{fig:ITO-aluminum}
\end{figure}

\begin{figure}[H]
    \centering
    \includegraphics[width=0.8\linewidth]{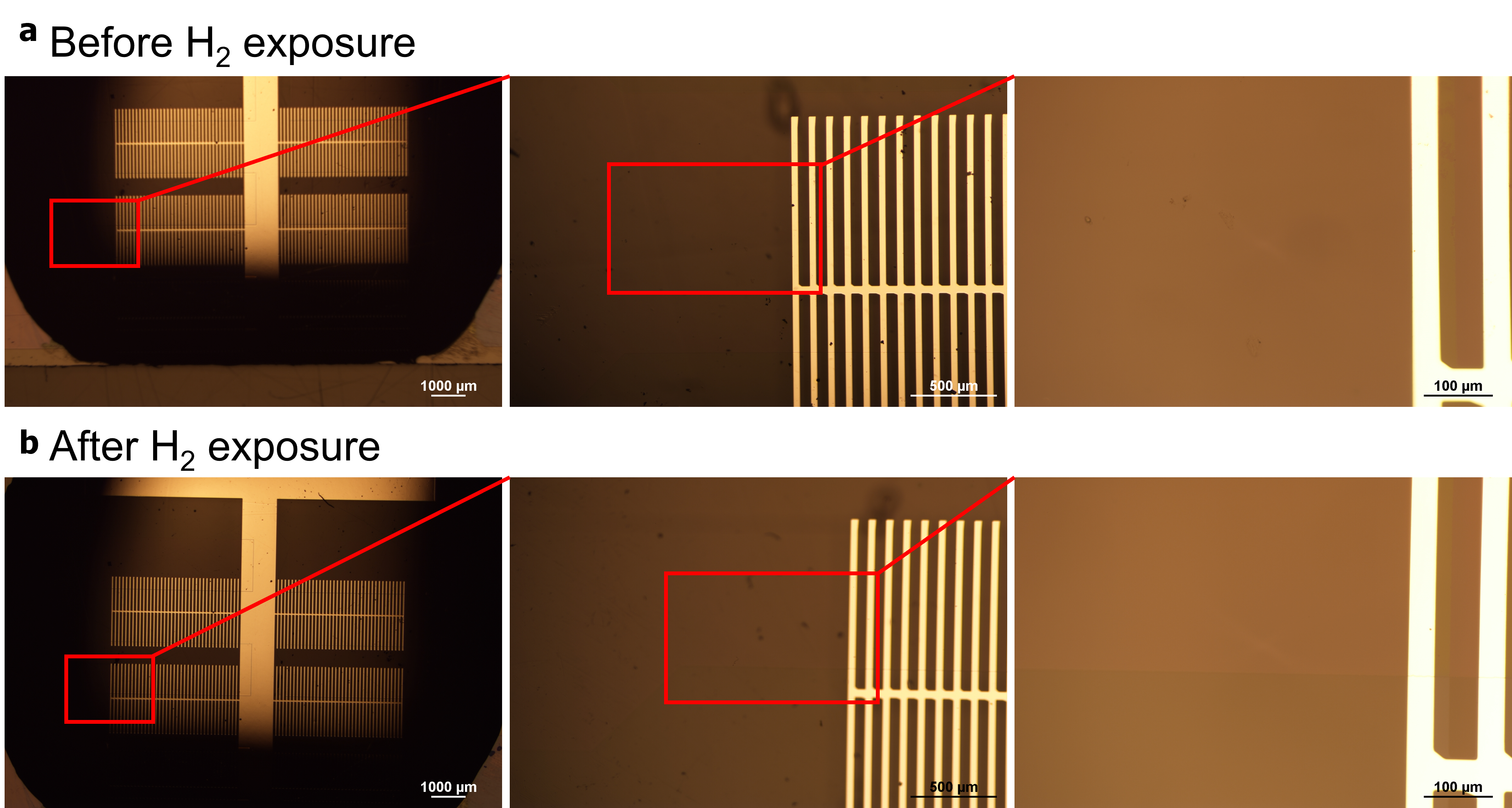}
    \caption{Microscopy images of the device \textbf{a} before, and \textbf{b} after hydrogen exposure for 24 hours. The images were taken at three different magnifications, increasing from left to right from 1x, 5x, and 20x, respectively. The area next to the electrodes, and for sure underneath, consists of the organic layers, which do not change in morphology under hydrogen exposure.}
    \label{fig:microscopy-before-after-H2}
\end{figure}
\FloatBarrier
\begin{figure}[H]
    \centering
    \includegraphics[width=1\linewidth]{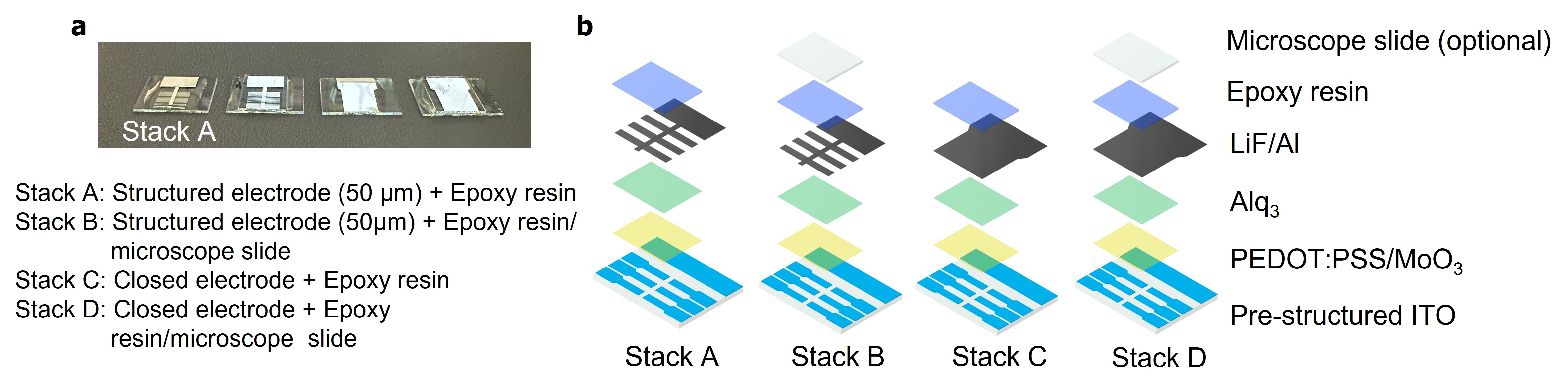}
    \caption{\textbf{a} Variation of encapsulation and electrode configuration. The underlying layers, namely $\mathrm{Alq_3}$, and the $\mathrm{PEDOT:PSS/MoO_3}$ layer were kept constant. \textbf{b} Scheme of the four different combinations of electrode and encapsulation.}
    \label{fig:encapsulation-sketch}
\end{figure}

\begin{figure}[H]
    \centering
    \includegraphics[width=1\linewidth]{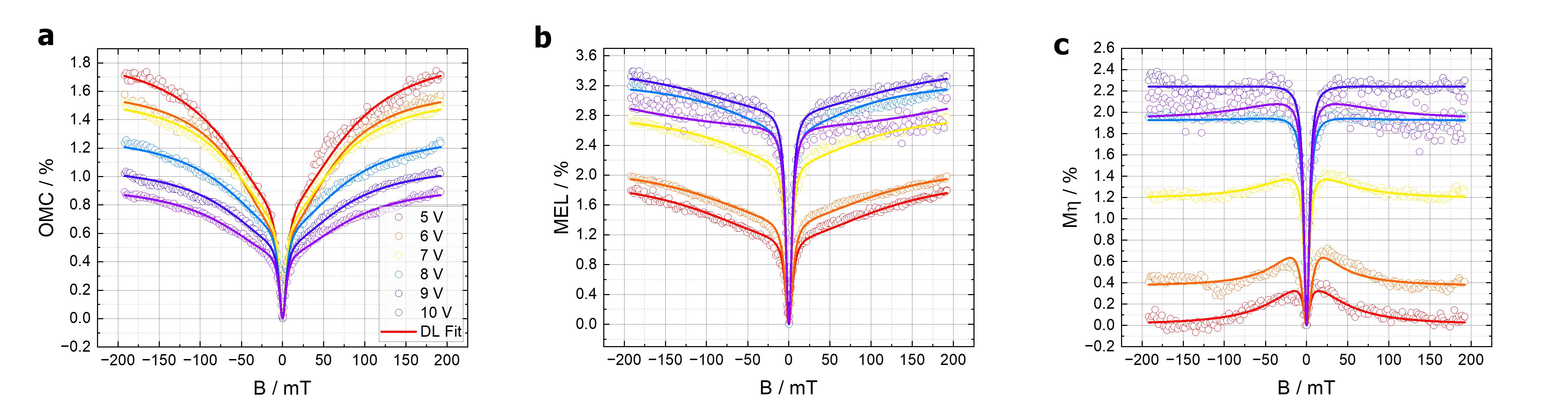}
    \caption{In \textbf{a-c} the magnetic field effects are shown depending on bias voltage for $\mathrm{Alq_3}$, with the magneto conductivity (\textbf{a}), the magneto electroluminescence \textbf{b}, and the calculated magneto efficiency (\textbf{c}) }
    \label{fig:magnetic field effects}
\end{figure}

\FloatBarrier
According to our previous publication \cite{en17205158}, we proposed that the sensor response is influenced by an external magnetic field \cite{en17205158, morgenstern2025unlocking}. Several organic materials were reported to exhibit sensitivity to external magnetic fields, which is generally ascribed to modifications of charge-carrier transport arising from Zeeman splitting of spin-dependent energy levels \cite{ehrenfreund2012effects}. To monitor this, the organic magneto-conductance(OMC) and the organic magneto-electroluminescence (MEL) response were measured as a function of the applied bias voltage and the external magnetic field. OMC and MEL responses can be described as follows. 

\begin{equation}
    \mathrm{OMC}(B) = \frac{\eta(B)- \eta(\mathrm{B} = 0)}{\eta(\mathrm{B} = 0)},
\end{equation}
\begin{equation}
    \mathrm{MEL}(B) = \frac{\mathrm{EL}(B)- \mathrm{EL}(\mathrm{B} = 0)}{\mathrm{EL}(\mathrm{B} = 0)},
\end{equation}
with $\eta$ being the conductivity, EL the electroluminescence, and B the external magnetic field.\newline
Additionally, the magneto-efficiency ($\mathrm{M_\eta = MEL - OMC}$) was calculated from both signals. For more detailed information, we refer to our previous publication \cite{morgenstern2025unlocking}. The results are depicted in Figure \ref{fig:magnetic field effects}. 

\begin{figure}[H]
    \centering
    \includegraphics[width=0.6\linewidth]{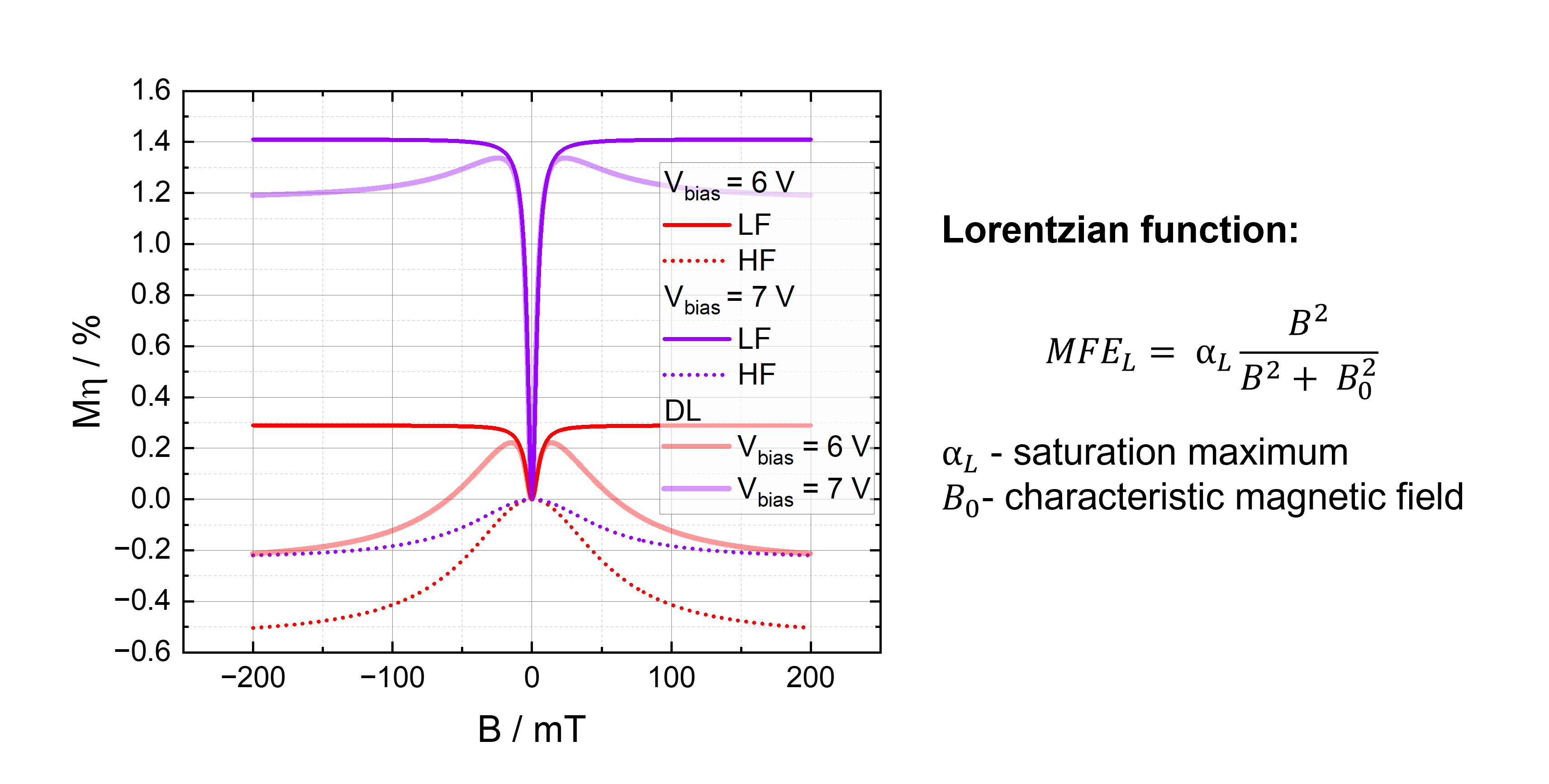}
    \caption{Magnetoefficiency depicted for a bias voltage of \SI{6}{V} and \SI{7}{V}. Additionally, the two Lorentzian contributions are shown, namely low-field (LF) and high-field (HF) effects. Increasing the bias voltage increases the contribution of the low-field effect, while the HF effect contribution is lowered. On the right-hand side, the equation of the Lorentzian function is shown, corresponding to a single term.}
    \label{fig:Lorentzian-Fit-magnetoefficiency}
\end{figure}

The curves were fitted with a combination of two Lorentzian fitting functions (cf. Figure \ref{fig:Lorentzian-Fit-magnetoefficiency}). For more details regarding the fitting procedure, we refer to Figure S5, where the two contributions, namely low- and high-field effect, are plotted exemplarily for a bias voltage of \SI{6}{V} and \SI{7}{V}. 

$\mathrm{Alq_3}$ shows a typical hyperfine-induced spin-mixing between the polaron-pair species, indicated by the narrow curve shape in the low-field regime ($B = \pm 20 \, \mathrm{mT}$), represented by the first Lorentzian fit term. Polaron pairs (PP) are loosely bound electron–hole pairs that differ from excitons in their binding energy according to the extended electron–hole separation distance. While excitons are tightly bound with binding energies in the range of a few 100 meV, PP are more weakly bound, as their spatial separation is larger \cite{bassler2011charge}. PP can be either in a singlet or triplet spin configuration, depending on their mutual polaron spin configuration \cite{kohler2015electronic}. If the effective decay rates for triplet and singlet PP are not equal, they are highly sensitive towards disturbance, \textit{e.g.} the application of an external magnetic field, resulting in a magnetic field dependence of the EL and $\eta$ \cite{ehrenfreund2012effects}. 

The second Lorentzian term was used to fit the outer part of the curve, namely the high-field regime ($B > 20\ \mathrm{mT}$). As evident by the decrease in $M_\eta$ response at higher magnetic fields, especially for the lower bias voltages of 5 and 6 V, a triplet-triplet annihilation (TTA) mechanism was identified \cite{morgenstern2025unlocking, wang2024understanding}. As previously reported by Cölle \textit{et al.} \cite{colle2004triplet}, $\mathrm{Alq_3}$ exhibits phosphorescence and delayed fluorescence caused by TTA. Nevertheless, in the operating regime of the sensor, at 7 V bias voltage, the TTA plays only a minor role, indicated by the small contribution of the second Lorentzian term and the line shape of the MEL response, which does slightly decrease at higher magnetic fields (cf. Figure S5). Hence, the dominant mechanism in these devices is attributed to hyperfine interactions involving polaron-pair species, consistent with previous observations for $\mathrm{Alq_3}$ \cite{mondal2023degradation, liu2009magnetic}.

Nguyen \textit{et al.} \cite{nguyen2007magnetoresistance} claimed that the hyperfine coupling induced by the proton's nuclear spin is a prerequisite for the observation of organic magnetic field effects.  As the hyperfine interaction is also dominated by the presence of hydrogen atoms, evident by isotopic studies, where deuterated and protonated molecules were compared \cite{nguyen2012isotope}, we initially expected the sensing effect to be based on the MFE \cite{en17205158}. Thus, the hyperfine interaction changes upon the hydrogen exposure, as more hydrogen is present. Surprisingly, the sensor signal was also responsive to hydrogen exposure without an external magnetic field, but, as discussed in the manuscript, an external magnetic field had an influence on the rise and fall time of the sensor response

\begin{figure} [H]
    \centering
    \includegraphics[width=0.7\linewidth]{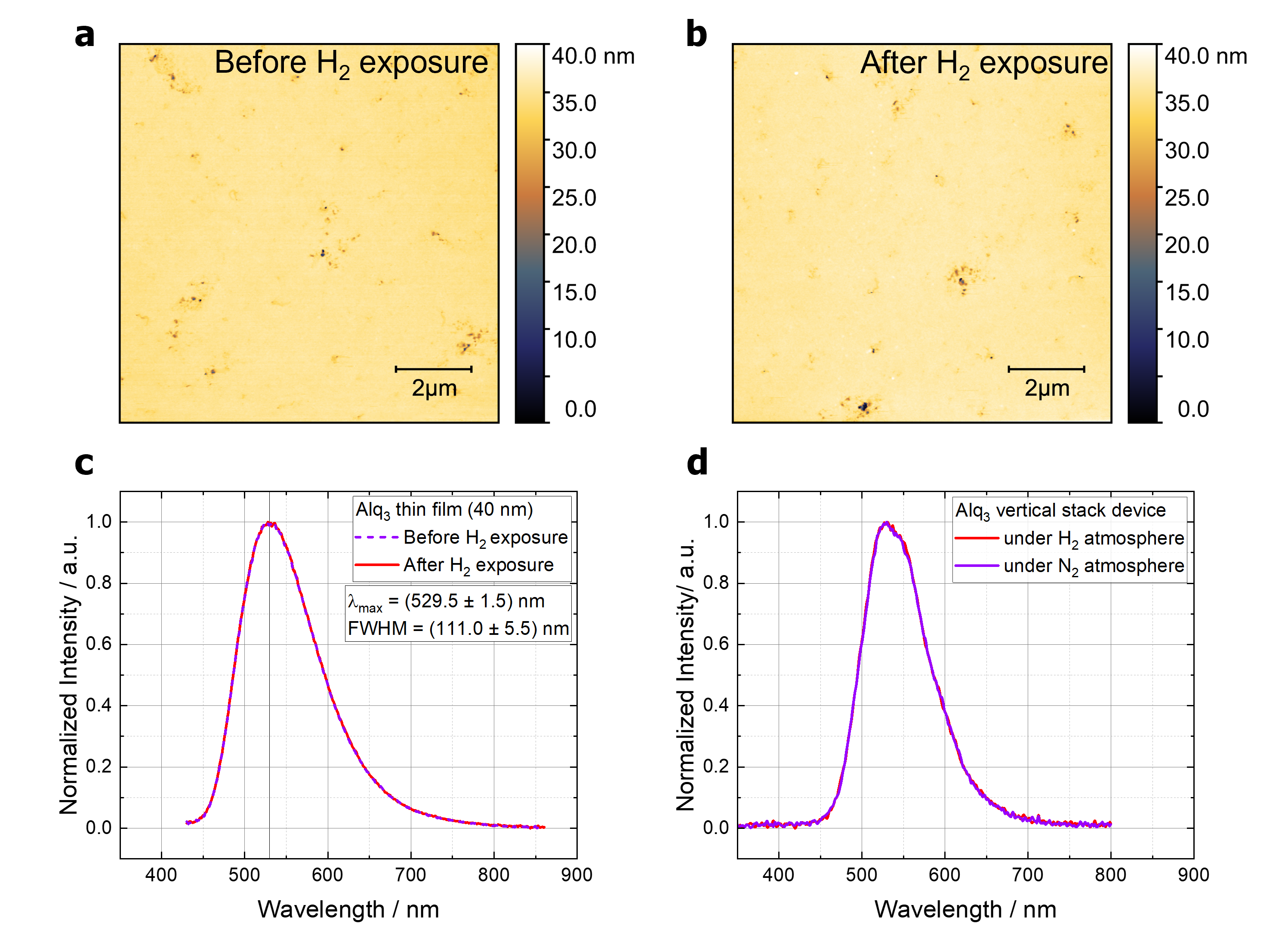}
    \caption{An $\mathrm{Alq_3}$ thin film was produced the same way the film is produced in the layer stack of the sensor. The film was exposed to hydrogen of $100\mathrm{\% \, vol}$ for two hours. The film was examined by AFM before \textbf{a} and after \textbf{b} hydrogen exposure. Additionally, the photoluminescence spectra were recorded before and after the exposure to hydrogen gas \textbf{c}. In \textbf{d} the EL spectra are shown under $\mathrm{N_2}$ and $\mathrm{H_2}$ exposure.}
    \label{fig:AFM-PL-Alq3}
\end{figure}

\begin{figure}[H]
    \centering
    \includegraphics[width=1\linewidth]{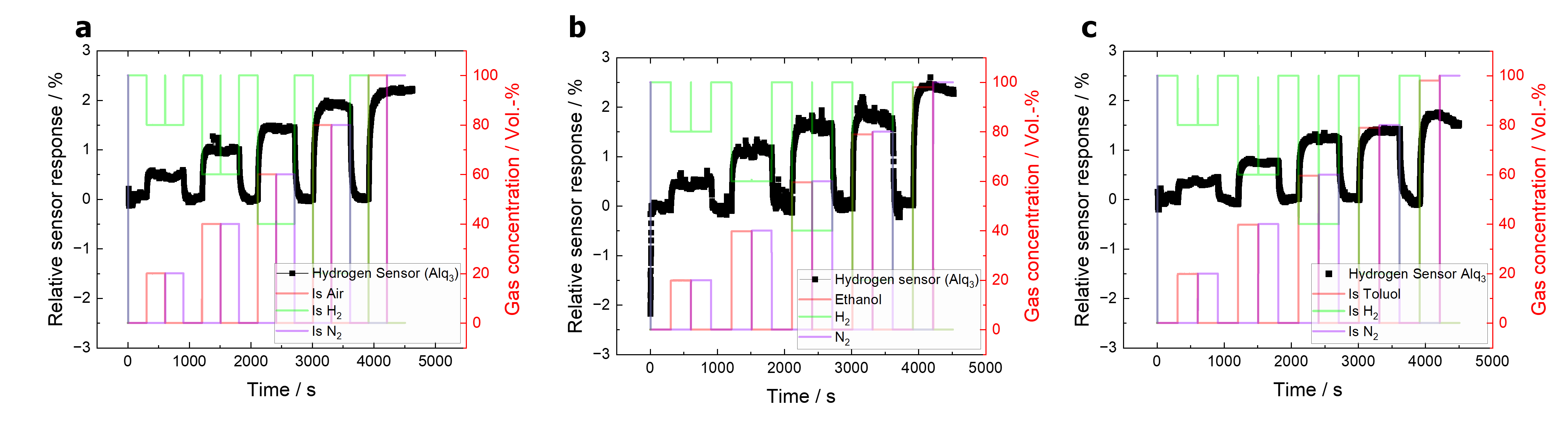}
    \caption{Raw data for the determination of cross-sensitivity towards \textbf{a} dry air, \textbf{b} ethanol, and \textbf{c} toluene. The gas under investigation was alternatively turned on with the carrier gas nitrogen. Additionally, the hydrogen gas concentration was constantly decreased from $100\mathrm{\% \, vol}$ to $0\mathrm{\% \, vol}$ in $20\mathrm{\% \, vol}$ steps.}
    \label{fig:cross-sensitivity-raw data}
\end{figure}

\FloatBarrier
\clearpage
\newpage
\bibliographystyle{MSP}
\bibliography{sn-bibliography}% common bib file
\end{document}